\documentclass[iop]{emulateapj}
\usepackage{psfig,amsfonts,amsmath,graphicx,apjfonts,subfigure}
\def\ra#1#2#3{#1$^{\rm h}$#2$
^{\rm m}$#3$^{\rm s}$}
\def\dec#1#2#3{$#1^\circ#2'#3''$}
\def\nod{\nodata}
\def\swift{{\it Swift}}
\def\chandra{{\it Chandra}}

\def\har{1}
\def\war{2}
\def\lei{3}
\def\psu{4}
\def\cal{5}
\def\ber{6}
\def\car{7}
\def\aim{8}
\def\got{9}

\shorttitle{Short GRB Environments}
\shortauthors{Fong et al.}

\begin{document}

\title{Demographics of the Galaxies Hosting Short-duration Gamma-Ray Bursts}

\author{
W.~Fong\altaffilmark{\har},
E.~Berger\altaffilmark{\har},
R.~Chornock\altaffilmark{\har},
R.~Margutti\altaffilmark{\har},
A.~J.~Levan\altaffilmark{\war},
N.~R.~Tanvir\altaffilmark{\lei},
R.~L.~Tunnicliffe\altaffilmark{\war},
I.~Czekala\altaffilmark{\har},
D.~B.~Fox\altaffilmark{\psu},
D.~A.~Perley\altaffilmark{\cal},
S.~B.~Cenko\altaffilmark{\ber},
B.~A.~Zauderer\altaffilmark{\har},
T.~Laskar\altaffilmark{\har},
S.~E.~Persson\altaffilmark{\car},
A.~J.~Monson\altaffilmark{\car},
D.~D.~Kelson\altaffilmark{\car},
C.~Birk\altaffilmark{\car},
D.~Murphy\altaffilmark{\car},
M.~Servillat\altaffilmark{\har}$^{,}$\altaffilmark{\aim},
G.~Anglada\altaffilmark{\got}
}

\altaffiltext{1}{Harvard-Smithsonian Center for Astrophysics, 60
Garden Street, Cambridge, MA 02138, USA}

\altaffiltext{2}{Department of Physics, University of Warwick, Coventry CV4 7AL,
UK}

\altaffiltext{3}{Department of Physics and Astronomy, University of Leicester, University Road, Leicester LE1 7RH, UK}

\altaffiltext{4}{Department of Astronomy and Astrophysics, 525
Davey Laboratory, Pennsylvania State University, University Park, PA
16802, USA}

\altaffiltext{5}{Cahill Center for Astronomy and Astrophysics, Room 232,
California Institute of Technology
Pasadena, CA 91125, USA}

\altaffiltext{6}{Department of Astronomy, University of California, Berkeley, CA 94720-3411, USA}

\altaffiltext{7}{Observatories of the Carnegie Institution of Washington, 813 Santa Barbara Street, Pasadena, CA 91101, USA}

\altaffiltext{8}{Laboratoire AIM (CEA/DSM/IRFU/SAp, CNRS, Universite Paris Diderot),
CEA Saclay, Bat. 709, 91191 Gif-sur-Yvette, France}

\altaffiltext{9}{Universit$\ddot{\rm a}$t G$\ddot{\rm o}$ttingen, Institut f$\ddot{\rm u}$r Astrophysik, Friedrich-Hund-Platz 1, 37077 G$\ddot{\rm o}$ttingen, Germany}

\begin{abstract}

We present observations of the afterglows and host galaxies of three
short-duration gamma-ray bursts (GRBs): 100625A, 101219A and
110112A. We find that GRB\,100625A occurred in a $z=0.452$
early-type galaxy with a stellar mass of $\approx 4.6 \times
10^{9}\,M_{\odot}$ and a stellar population age of $\approx 0.7$ Gyr,
and GRB\,101219A originated in a star-forming galaxy at $z=0.718$ with
a stellar mass of $\approx 1.4 \times 10^{9}\,M_{\odot}$, a star
formation rate of $\approx 16$ $M_{\odot}$ yr$^{-1}$, and a stellar
population age of $\approx 50$ Myr. We also report the discovery of
the optical afterglow of GRB 110112A, which lacks a coincident host
galaxy to $i\gtrsim 26$ mag and we cannot conclusively identify any
field galaxy as a possible host. From afterglow modeling, the bursts
have inferred circumburst densities of $\approx 10^{-4}-1$
cm$^{-3}$, and isotropic-equivalent gamma-ray and kinetic energies of
$\approx 10^{50}-10^{51}$ erg. These three events highlight the
diversity of galactic environments that host short GRBs. To quantify
this diversity, we use the sample of $36$ \swift\ short GRBs with
robust associations to an environment ($\sim 1/2$ of $68$ short bursts
detected by \swift\ to May 2012) and classify bursts originating from
four types of environments: late-type ($\approx 50\%$), early-type ($
\approx 15\%$), inconclusive ($\approx 20\%$), and ``host-less''
(lacking a coincident host galaxy to limits of $\gtrsim 26$ mag;
$\approx 15\%$). To find likely ranges for the true late- and
early-type fractions, we assign each of the host-less bursts to either
the late- or early-type category using probabilistic arguments, and
consider the scenario that all hosts in the inconclusive category are
early-type galaxies to set an upper bound on the early-type
fraction. We calculate most likely ranges for the late- and early-type
fractions of $\approx 60-80\%$ and $\approx 20-40\%$, respectively. We
find no clear trend between gamma-ray duration and host type. We
also find no change to the fractions when excluding events recently
claimed as possible contaminants from the long GRB/collapsar
population. Our reported demographics are consistent with a short GRB rate
driven by both stellar mass and star formation.

\end{abstract}


\section{Introduction}

Observations of the galactic environments of cosmic explosions provide
invaluable insight into their underlying progenitor populations. For
example, Type Ia supernovae (SNe) originate in both star-forming and
elliptical galaxies \citep{ot79,vlf05,mdp+05,lcl+11} consistent with
an evolved progenitor and an event rate that traces both stellar mass
and star formation \citep{slp+06}. In contrast, SNe of types II and
Ib/c are found to occur only in spiral and irregular galaxies,
indicating that these events result from the core-collapse of young,
massive stars \citep{vlf05,hpm+08,lcl+11} and a rate tracing recent star
formation \citep{kk12,ahj+12}.

In the case of long-duration gamma-ray bursts, (GRBs; $T_{90}\gtrsim
2$ s; \citealt{kmf+93}) the link to star-forming host galaxies
helped to establish that their progenitors are massive stars
\citep{dkb+98,ldm+03,fls+06,wbp+07}. Furthermore, a decade of
concerted efforts to characterize the stellar populations of long GRB
hosts revealed young stellar population ages of $\lesssim 0.2$ Gyr, a
mean stellar mass of $\approx 2 \times 10^{9}\,M_{\odot}$, and
inferred UV/optical star formation rates (SFR) of $\approx 1-50\,M_{\odot}$
yr$^{-1}$ \citep{chg+04,sgl09,lb10,lbc11}. In addition, the spatial
locations of long GRBs with respect to their host galaxy centers (with
a mean of $\sim 1$ half-light radius; \citealt{bkd02}) and their
concentration in bright UV regions of their hosts \citep{fls+06}
provided a direct association between long GRBs and star formation.

In contrast, the origin of short GRBs ($T_{90} \lesssim 2$ s) is less
clear, as the first few afterglow discoveries led to associations with
both elliptical \citep{bpc+05,cdg+05,gso+05,hsg+05,bpp+06} and star-forming
\citep{ffp+05,hwf+05,sbk+06,gbp+06,bgc+06} host galaxies, demonstrating that at least some
short GRBs originate from older stellar populations. Studies primarily
focused on the sample of bursts with sub-arcsecond localization have
shown the population of hosts to be dominated by late-type galaxies,
albeit with lower specific SFRs, higher luminosities, and higher
metallicities than the star-forming hosts of long GRBs \citep{ber09}.
Modeling of the spectral energy distributions of short GRB host
galaxies has led to a broad range of inferred ages, $\tau \approx
0.03-4.4$ Gyr, and an average stellar mass of $\approx 2 \times
10^{10}\,M_{\odot}$ \citep{lb10}. A detailed analysis of their
sub-galactic environments through {\it Hubble Space Telescope}
observations has demonstrated that on average, short GRBs have
offsets from their hosts of $\approx 5$ kpc \citep{fbf10}, while a
growing subset which lack coincident hosts may have offsets of
$\gtrsim 30$ kpc \citep{ber10}. Finally, an examination of short GRB
locations with respect to their host light distributions revealed that
they under-represent their host UV/optical light \citep{fbf10}. These
results are consistent with theoretical expectations for NS-NS/NS-BH
mergers \citep{elp+89,npp92}, with potential minor contribution from
other proposed progenitors, such as the accretion-induced-collapse of
a WD or NS \citep{qwc+98,lwc+06,mqt08} or magnetar flares
\citep{lwc+06,clw+08}.

However, the majority of short GRB host galaxy studies published thus far
primarily concentrate on bursts with sub-arcsecond localization from
optical afterglows. While these events have the most unambiguous
associations with host galaxies, the fraction is only $\sim 1/3$
($23/68$ to May 2012) of all short GRBs detected by the \swift\
satellite \citep{ggg+04}. The faintness of their optical afterglows
($\approx 23$ mag at $\sim 10$ hr after the burst; \citealt{ber10}) is
likely attributed to a combination of a low energy scale \citep{pkn01} and
circumburst densities. Therefore, if there exist correlations between
these basic properties and host galaxy type, the selection by optical
afterglows may affect the relative rates of short GRBs detected in
early- and late-type host galaxies. An alternative route to
sub-arcsecond localization is through the X-ray detection of an
afterglow, which does not necessarily depend on circumburst density
\citep{gs02} with \chandra; however, only two such cases have been
reported thus far \citep{fbm+12,mbf+12,sta+12}.

Demographics which accurately represent the bulk of the short GRB
population are imperative in understanding the link to the
progenitors. In particular, the late-to-early-type host galaxy ratio
will inform whether stellar mass or SFR drives the
short GRB rate \citep{lb10}, and will help to constrain the delay time
distribution \citep{zr07}. Furthermore, a recent study based on
$\gamma$-ray properties (spectral hardness and duration) claims that
there is a non-negligible fraction of contaminants from collapsars in
the \swift\ short GRB population \citep{bnp+12}. Thus, an examination
of how this fraction affects the environment demographics will aid in
assessing the true contamination.

Fortunately, the detection of X-ray afterglows with \swift/XRT
\citep{ggg+04,bhn+05} enables positions with $\sim$few arcsecond
precision in $\approx 60\%$ ($40/68$) of all \swift\ short GRBs. In
the majority of such cases, these XRT positions coupled with dedicated
optical/NIR searches for host galaxies have provided meaningful
associations to a galactic environment\footnotemark[10]\footnotetext[10]{The
large majority of the remaining $\approx 40\%$ of \swift\ short GRBs
lack afterglow follow-up due to observing constraints unrelated to the
burst properties; see \S5.}. While such bursts with XRT positions have
been studied as single events
(e.g. \citealt{gso+05,bpp+06,bpc+07,pmm+12}), the entire sample has
not been studied in detail alongside bursts with sub-arcsecond
localization.

To this end, we present here X-ray and optical/NIR observations of the
afterglows and environments of three short
GRBs\footnotemark[11]\footnotetext[11]{We present observations of two
additional short GRBs, 100628A and 100702A, both with published
\swift/XRT localizations (see Appendix). We show that the XRT
afterglow of GRB\,100628A is of low significance, while the XRT
position of GRB\,100702A is contaminated, preventing an unambiguous
association with a host galaxy.} localized by \swift/XRT, which
highlight the diversity of their galactic environments: GRBs\,100625A,
101219A, and 110112A. We also present the discovery of the optical
afterglow of GRB\,110112A. While GRBs\,100625A and 101219A have robust
associations with host galaxies, GRB\,110112A lacks a coincident host
to deep optical limits. We describe the X-ray, optical and NIR
observations for these three events (\S2), present their energy scales
and circumburst densities inferred from afterglow modeling (\S3), and
host galaxy stellar population ages, masses and SFRs extracted from spectroscopy
and broad-band SEDs (\S4). We discuss the stellar population
characteristics of these three host galaxies compared to previous short GRB
hosts (\S5). Putting these bursts into the context, we undertake the
first comprehensive study of host galaxy demographics of both
sub-arcsecond localized and XRT-localized bursts, by investigating the
late- and early-type host galaxy fractions for the bulk of the short
GRB population, and compare host galaxy type to $\gamma$-ray
properties (\S6).

Unless otherwise noted, all magnitudes are in the AB system and are
corrected for Galactic extinction in the direction of the burst
\citep{sfd98,sf11}, and uncertainties correspond to $1\sigma$
confidence. We employ a standard $\Lambda$CDM cosmology with
$\Omega_M=0.27$, $\Omega_\Lambda=0.73$, and $H_0=71$ km s$^{-1}$
Mpc$^{-1}$.

\section{Observations}

\begin{deluxetable*}{lccccccc}
\tabletypesize{\scriptsize}
\tablecolumns{8}
\tabcolsep0.0in\footnotesize
\tablewidth{0pc}
\tablecaption{Short GRB Properties
\label{tab:info}}
\tablehead {
\colhead {GRB}           &
\colhead {R.A.}          &
\colhead {Decl}          &
\colhead {Uncert.}       &
\colhead {$z$}           &
\colhead {T$_{90}$ ($15-350$ keV)}      &
\colhead {$f_{\gamma}$ ($15-150$ keV)}  &
\colhead {References}    \\
\colhead {}              &
\colhead {(J2000)}       &
\colhead {(J2000)}       &
\colhead {($''$)}        &
\colhead {}              &
\colhead {(s)}           & 
\colhead {(erg cm$^{-2}$)} &
\colhead {}         
}
\startdata
GRB\,100625A & \ra{01}{03}{10.91} & \dec{-39}{05}{18.4} & $1.8$ & $0.452$ & $0.33 \pm 0.03$ & $(2.3 \pm 0.2) \times 10^{-7}$ & $1$ \\
GRB\,101219A & \ra{04}{58}{20.49} & \dec{-02}{32}{23.0} & $1.7$ & $0.718$ & $0.6 \pm 0.2$   & $(4.6 \pm 0.3) \times 10^{-7}$ & $2$ \\
GRB\,110112A & \ra{21}{59}{43.85} & \dec{+26}{27}{23.9} & $0.14$ & \nod    & $0.5 \pm 0.1$   & $(3.0 \pm 0.9) \times 10^{-8}$ & $3$, This work            
\enddata
\tablecomments{{\bf References}: (1) \citealt{gcnr289}; (2) \citealt{gcn11467}; (3) \citealt{gcn11557} 
}
\end{deluxetable*}

\subsection{GRB\,100625A}

GRB\,100625A was detected by three $\gamma$-ray satellites on 2010
June 25.773~UT: the Burst Alert Telescope (BAT) on-board the \swift\
satellite \citep{ggg+04,gcn10884}, {\it Konus-Wind} \citep{gcn10890}
and the Gamma-Ray Burst Monitor (GBM) on-board {\it Fermi}
\citep{gcn10912}. BAT localized the burst to a ground-calculated
position of RA=\ra{01}{03}{11.1}, Dec=$-$\dec{39}{05}{29} (J2000) with
an uncertainty of $1.0'$ radius ($90\%$ containment;
\citealt{gcnr289}), and the burst consisted of two pulses with a total
duration of $T_{90}=0.33 \pm 0.03$s ($15-350$ keV) and a fluence of
$f_{\gamma} = (2.3 \pm 0.2) \times 10^{-7}$ erg cm$^{-2}$ ($15-150$
keV; \citealt{gcnr289}). {\it Fermi}/GBM observations determined
$E_{\rm peak} = 509^{+77}_{-61}$~keV and $f_\gamma=(1.32 \pm 0.05)
\times 10^{-6}$~erg~cm$^{-2}$ ($8-1000$~keV; \citealt{gcn10912}),
while {\it Konus-Wind} observations determined $E_{\rm
peak}=418^{+128}_{-78}$~keV and $f_\gamma=(8.3 \pm 1.5) \times
10^{-7}$~erg~cm$^{-2}$ ($20-2000$~keV; \citealt{gcn10890}). Based on the short
duration and high $E_{\rm peak}$, GRB\,100625A can be classified as a short,
hard burst. The $\gamma$-ray properties are listed in
Table~\ref{tab:info}.

\subsubsection{X-ray Observations}
\label{sec:100625prompt}

The X-ray Telescope (XRT) on-board \swift\ began observing the field
at $\delta t=43$ s ($\delta t$ is the time after the BAT trigger) and
detected a fading, uncatalogued X-ray source at RA=\ra{01}{03}{10.91}
and Dec=\dec{-39}{05}{18.4} with a final positional accuracy of
$1.8''$ radius ($90\%$; \citealt{gtb+07,ebp+09,gcnr289};
Table~\ref{tab:info}).

\begin{figure}
\centering
\includegraphics[angle=0,width=3.2in]{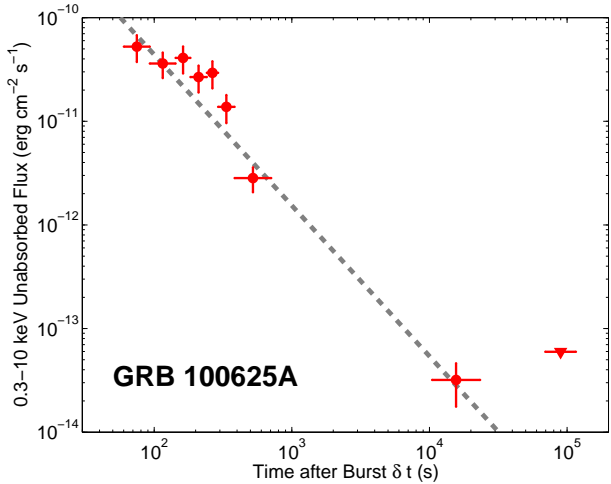}
\caption{\swift/XRT light curve of GRB\,100625A. The
 triangle is a $3\sigma$ upper limit. The entire light curve is
best fit with a power law characterized by $\alpha_X=-1.45 \pm 0.08$
(grey dashed line).
\label{fig:100625_xraylc}}
\end{figure}

We analyze the XRT data using HEASOFT (v.6.11) and relevant
calibration files. We apply standard filtering and screening criteria,
and generate a count rate light curve following the prescriptions from
\citet{mgg+10} and \citet{mzb+12}. Our re-binning scheme ensures a
minimum signal-to-noise ratio of $S/N=4$ for each temporal bin. To
extract a spectrum, we use Cash statistics and fit the XRT data with
an absorbed power law model ($tbabs \times ztbabs \times pow$ within
the XSPEC routine) characterized by photon index, $\Gamma$, and
intrinsic neutral hydrogen absorption column density,
$N_{\rm{H,int}}$, in excess of the Galactic column density in the
direction of the burst, $N_{{\rm H,MW}}=2.1\times
10^{20}\,\rm{cm^{-2}}$ (typical uncertainty of $\sim 10\%$;
\citealt{kbh+05,wlb11}). We utilize the entire PC data set ($\delta
t=60-10^{5}$ s), where there is no evidence for spectral
evolution. Our best-fit spectrum (C-stat$_\nu=0.92$ for $95$ d.o.f.)
is characterized by $\Gamma=2.5 \pm 0.2$ and $N_{\rm H,int} \lesssim
1.7 \times 10^{21}$ cm$^{-2}$ ($3\sigma$) at $z=0.452$ (see
\S\ref{sec:100625spec} for the redshift determination). Our
best-fit parameters are consistent with the automatic spectrum fit
produced by \citet{gcn10888}. Applying these parameters to the data,
we calculate the count rate-to-flux conversion
factors, and hence the unabsorbed fluxes
(Figure~\ref{fig:100625_xraylc}).

To quantify the decay rate, we utilize $\chi^2$-minimization to fit a
power law to the data in the form $F_X(t) \propto t^{\alpha_X}$, with
$\alpha_X$ as the free parameter. The entire XRT light curve ($\delta
t \approx 80-10^5$ s, PC mode) is best fit with a single power law
with index $\alpha_X=-1.45 \pm 0.08$ ($\chi^2_{\nu}=2.1$ for 7 d.o.f.;
Figure~\ref{fig:100625_xraylc}).

\subsubsection{Optical/NIR Observations and Afterglow Limits}
\label{sec:100625obs}

\begin{figure*}
\centering
\includegraphics*[angle=0,width=6.4in]{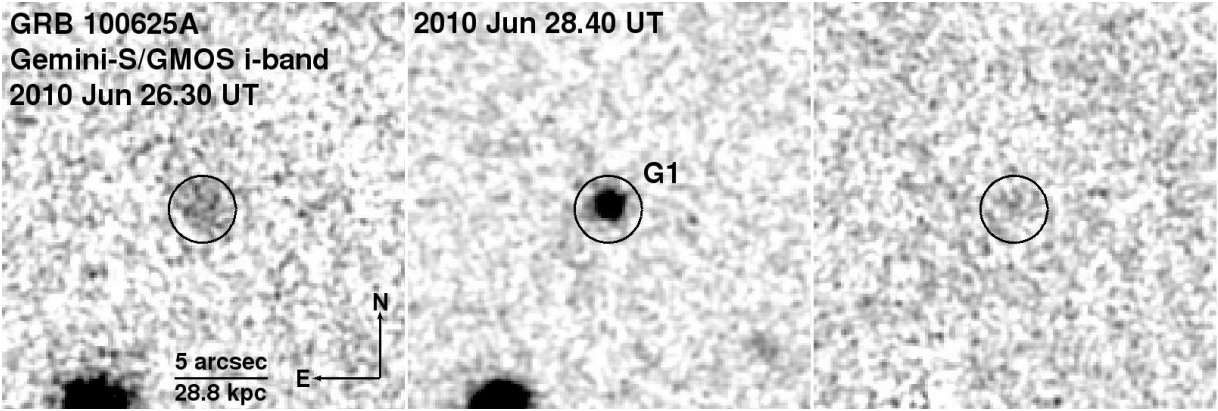}
\caption{Gemini-South/GMOS $i$-band observations of GRB\,100625A. The
XRT error circle has a radius of $1.8''$ ($90\%$ containment;
black). Images are smoothed with a $2$-pixel Gaussian. {\it Left:}
$\delta t=0.53$~d in poor seeing conditions ($\theta_{\rm
FWHM}=1.9''$) with a faint host detection. {\it Center:} $\delta
t=2.63$~d with $0.9''$ seeing. {\it Right:} Digital image subtraction
of the two epochs reveals no afterglow to a $3\sigma$ limit of $i
\gtrsim 22.7$ mag. The host galaxy is marked as G1.
\label{fig:100625}}
\end{figure*}

The UV-Optical Telescope (UVOT) on-board \swift\ commenced
observations at $\delta t=56$ s but no corresponding source was found
within the XRT position. The $3\sigma$ limit over
$\delta t \approx 87-1.2 \times 10^{4}$ s in the $white$ filter, which
transmits over $\lambda=1600$-$7000$~\AA\ \citep{pbp+08}, is $\gtrsim
22.6$ mag (not corrected for Galactic extinction;
\citealt{gcnr289}). Rapid ground-based follow-up in the optical and
NIR provided early limits on the afterglow of $I \gtrsim 22.8$ mag at
$\delta t \approx 17$ min \citep{gcn10885} and $J \gtrsim 19.4$ mag at
$\delta t \approx 8.6$ hr \citep{gcn10889}. GROND observations at
$\delta t \approx 12.2$ hr place limits of $g \gtrsim 23.6$ mag, and
$riz \gtrsim 23$ mag \citep{nkg+12}.

We obtained optical observations of GRB\,100625A with the Gemini
Multi-Object Spectrograph (GMOS) mounted on the Gemini-South $8$-m
telescope, starting at $\delta t=12.4$ hr in the $riz$ filters in poor
seeing conditions (Table~\ref{tab:obs}). We analyze the data using the
IRAF {\tt gemini} package, and detect a single source within the
enhanced XRT error circle in all three filters. To assess any
potential fading of the source, we obtained a second set of
observations at $\delta t \approx 2.6$~d, where the source is clearly
extended. Digital image subtraction using the ISIS software package
\citep{ala00} shows no residuals in all three filters
(Figure~\ref{fig:100625}). We therefore place $3\sigma$ limits of $r
\gtrsim 22.6$ mag, $i \gtrsim 22.7$ mag and $z \gtrsim 22.8$ mag on
the optical afterglow at $\delta t \approx 12.7$ hr
(Table~\ref{tab:obs}). The GMOS zeropoints are
determined by sources in common with late-time IMACS observations (see
below), which are calibrated to a standard star field at a similar
airmass. Our limits match the GROND limits reported at
$\delta t \approx 12.2$ hr \citep{nkg+12}. 

In addition, we obtained two epochs of $J$-band observations with the
Persson's Auxilliary Nasmyth Infrared Camera (PANIC) mounted on the
$6.5$-m Magellan/Baade telescope at $\delta t \approx 1.6$ and
$6.6$~d. We analyze the data using standard procedures in
IRAF. Digital image subtraction shows no evidence for fading, with a
$3\sigma$ limit of $J \gtrsim 23.9$ mag (photometrically tied to the
2MASS catalog and converted to the AB system) at $\delta t \approx
1.6$~d (Table~\ref{tab:obs}).

We obtained late-time $griz$ observations of the the field of
GRB\,100625A with the Inamori Magellan Areal Camera and Spectrograph
(IMACS) mounted on Magellan/Baade starting on 2010 November
14.11~UT. We also obtained $K_s$-band observations with the FourStar
Infrared Camera mounted on Magellan/Baade on 2011 December 07.16~UT
(Table~\ref{tab:obs}). The $griz$ zeropoints are calculated using a
standard star field at a similar airmass, while the $K_s$-band zeropoint is
determined from point sources in common with 2MASS. Our afterglow
limit and host galaxy photometry are summarized in
Table~\ref{tab:obs}.

We obtained a spectrum of the putative host galaxy with the Low
Dispersion Survey Spectrograph~3 (LDSS3) mounted on the 6.5-m
Magellan/Clay telescope on 2011 October 21.27~UT. A dithered pair of
$2700$~s exposures was obtained with the VPH-ALL grating, which has a
wavelength coverage of $4000-10000$ \AA\ and a spectral resolution of
$\approx 8$ \AA. We used standard tasks in IRAF for data reduction,
HeNeAr arc lamps for wavelength calibration, and observations of the
smooth-spectrum standard star EG131 for flux calibration. We discuss
the spectral features and redshift determination in
\S\ref{sec:100625spec}.

\tabletypesize{\scriptsize}
\begin{deluxetable*}{lccccccccccc}
\tablecolumns{12}
\tablewidth{0pc}
\tablecaption{Log of Optical/NIR Afterglow and Host Galaxy Photometry
\label{tab:obs}}
\tablehead {
\colhead {GRB}                 &
\colhead {Date}                &
\colhead {$\delta t$}          &
\colhead {Telescope}           &
\colhead {Instrument}          &
\colhead {Filter}              &
\colhead {Exposures}           &
\colhead {$\theta_{FWHM}$}     &
\colhead {Afterglow$^a$}           &
\colhead {$F_{\nu}^a$}           &
\colhead {Host$^a$}            &
\colhead {$A_{\rm \lambda,MW}$}     \\
\colhead {}                    &
\colhead {(UT)}                &
\colhead {(d)}                 &
\colhead {}                    &
\colhead {}                    &
\colhead {}                    &
\colhead {(s)}                 &
\colhead {($''$)}              &
\colhead {(AB mag)}            &
\colhead {($\mu$Jy)}        &
\colhead {(AB mag)}            &
\colhead {(mag)}                                   
}
\startdata
GRB\,100625A & 2010 Jun 26.288 & 0.52  & Gemini-S & GMOS     & $r$ & $5\times 120$ & $2.31$ & $>22.6$  & $<3.3$ & $22.76 \pm 0.23$ & $0.027$ \\
             & 2010 Jun 26.301 & 0.53  & Gemini-S & GMOS     & $i$ & $3\times 120$ & $1.91$ & $>22.7$  & $<2.9$ & $22.10 \pm 0.15$ & $0.020$ \\
             & 2010 Jun 26.314 & 0.54  & Gemini-S & GMOS     & $z$ & $5\times 120$ & $1.95$ & $>22.8$  & $<2.8$ & $22.23 \pm 0.15$ & $0.015$ \\
             & 2010 Jun 27.392 & 1.62  & Magellan & PANIC    & $J$ & $35\times 60$ & $0.76$ & $>23.9$  & $<1.0$ & $21.48 \pm 0.05$ & $0.008$ \\
             & 2010 Jun 28.394 & 2.62  & Gemini-S & GMOS     & $r$ & $5\times 120$ & $1.10$ &          &        & $22.63 \pm 0.09$ & $0.027$ \\
             & 2010 Jun 28.404 & 2.63  & Gemini-S & GMOS     & $i$ & $5\times 120$ & $0.87$ &          &        & $22.14 \pm 0.04$ & $0.020$ \\
             & 2010 Jun 28.414 & 2.64  & Gemini-S & GMOS     & $z$ & $5\times 120$ & $0.95$ &          &        & $22.07 \pm 0.10$ & $0.015$ \\
             & 2010 Jul 02.398 & 6.63  & Magellan & PANIC    & $J$ & $18\times 180$& $0.53$ &          &        & $21.40 \pm 0.06$ & $0.008$ \\
             & 2010 Nov 14.114 & 141.3 & Magellan & IMACS    & $g$ & $2\times 420$ & $0.65$ &          &         & $23.87 \pm 0.19$ & $0.039$ \\
             & 2010 Nov 14.123 & 141.4 & Magellan & IMACS    & $i$ & $1\times 240$ & $0.47$ &          &         & $22.04 \pm 0.07$ & $0.020$ \\
             & 2010 Nov 14.196 & 141.4 & Magellan & IMACS    & $r$ & $1\times 360$ & $0.65$ &          &         & $22.59 \pm 0.13$ & $0.027$ \\
             & 2010 Nov 14.200 & 141.4 & Magellan & IMACS    & $z$ & $1\times 180$ & $0.52$ &          &         & $21.88 \pm 0.22$ & $0.015$ \\
             & 2011 Dec 07.16  & 529.4 & Magellan & FourStar & $K_s$ & $90\times 10$ & $0.55$ &        &         & $20.76 \pm 0.10$ & $0.008$ \\
\hline
GRB\,100702A & 2010 Jul 02.10 & 0.05 & Magellan & PANIC & $J$ & $9 \times 180$ & $0.53$ & $>23.3^{b}$ & $<1.70^{b}$ & $20.54 \pm 0.05$ / $21.30 \pm 0.07^{c}$ & $0.284$ \\
             & 2010 Jul 02.30 & 0.25 & Magellan & PANIC & $J$ & $9\times 180$ & $0.75$ &          &         & \nod$^{d}$ / $21.49 \pm 0.11$ & $0.284$ \\
             & 2011 Mar 06.37 & 247.3 & Magellan & IMACS & $i$ & $2\times 240$ & $0.83$ &          &         &   $>22.7$               & $0.679$ \\
\hline
GRB\,101219A & 2010 Dec 19.15 & 0.04 & Gemini-S & GMOS  & $i$ & $9\times 180$ & $0.66$ & $>24.9$ & $<0.40$ & $23.20 \pm 0.11$ & $0.097$ \\
             & 2010 Dec 19.16 & 0.05 & Magellan & FourStar & $J$ & $25 \times 60$ & $0.46$ & $>23.6$ & $<1.36$ & $22.43 \pm 0.13$ & $0.041$ \\
             & 2010 Dec 19.17 & 0.07 & Gemini-S & GMOS  & $r$ & $9\times 180$ & $0.80$ & $>24.9$ & $<0.40$ & $23.83 \pm 0.26$ & $0.131$ \\
             & 2010 Dec 19.20 & 0.09 & Gemini-S & GMOS  & $i$ & $9\times 180$ & $0.69$ & $>24.9$    & $<0.40$ & $23.40 \pm 0.09$ & $0.097$ \\
             & 2010 Dec 19.27 & 0.16 & Gemini-S & GMOS  & $r$ & $12\times 180$& $0.67$ & $>25.1$    & $<0.34$ & $23.73 \pm 0.10$ & $0.131$ \\
             & 2010 Dec 19.30 & 0.20 & Gemini-S & GMOS  & $i$ & $12\times 180$& $0.67$ &          &         & $23.19 \pm 0.08$ & $0.097$ \\
             & 2010 Dec 28.16 & 9.05 & Gemini-S & GMOS  & $r$ & $12\times 240$& $0.65$ &          &         & $23.95 \pm 0.05$ & $0.131$ \\
             & 2011 Jan 12.15 & 24.05 & Magellan & LDSS3 & $z$ & $6\times 180$ & $0.68$ &          &         & $23.22 \pm 0.16$ & $0.072$ \\
             & 2011 Jan 12.17 & 24.06 & Magellan & LDSS3 & $g$ & $5\times 180$ & $1.05$ &          &         & $24.57 \pm 0.08$ & $0.189$ \\
             & 2011 Dec 07.24 & 353.1 & Magellan & FourStar & $J$ & $15\times 60$ & $0.56$ &       &         & $22.11 \pm 0.19$ & $0.041$ \\
             & 2011 Dec 07.25 & 353.1 & Magellan & FourStar & $K_s$ & $90\times 10$ & $0.44$ &      &         & $21.55 \pm 0.21$ & $0.017$ \\
\hline
GRB\,110112A & 2011 Jan 12.18 & 0.64 & WHT      & ACAM  & $i$ & $2\times 300$ & $1.10$ & $22.77 \pm 0.29$  & $2.84 \pm 0.75$  & \nod   & $0.104$ \\
             & 2011 Jun 27.83 & 166.2 & Magellan & LDSS3 & $i$ & $5\times 240$ & $0.94$ & &         &  $>24.7$                & $0.104$ \\
             & 2011 Jun 27.83 & 166.3 & Magellan & LDSS3 & $r$ & $3\times 360$ & $1.11$ &  &         & $>25.5$               & $0.140$ \\
             & 2011 Jul 28.46 & 197.3 & Gemini-N & GMOS  & $i$ & $15\times 180$& $0.61$ &  &         & $>26.2$            & $0.104$ 
\enddata
\tablecomments{Limits correspond to a $3\sigma$ confidence level. \\
$^a$ These values are corrected for Galactic extinction \citep{sf11}. \\
$^b$ Only applies to approximately half of the error circle. \\
$^c$ Magnitudes for S1 and S4, respectively. \\
$^d$ S1 is blended with a neighboring bright star (Figure~\ref{fig:100702_J}) so we cannot perform photometry.
}
\end{deluxetable*}

\subsection{GRB\,101219A}

GRB\,101219A was detected by \swift/BAT \citep{gcn11461} and {\it
Konus-Wind} \citep{gcn11470} on 2010 December 19.105 UT. BAT localized
the burst at a ground-calculated position of RA=\ra{04}{58}{20.7} and
Dec=$-$\dec{02}{31}{37.1} with a $1.0'$ radius uncertainty ($90\%$
containment; \citealt{gcn11467}). The $\gamma$-ray light curve
exhibits a double-peaked structure with $T_{90} = 0.6 \pm 0.2$ s
($15-350$ keV) and $f_{\gamma} = (4.6 \pm 0.3) \times 10^{-7}$ erg
cm$^{-2}$ ($15-150$ keV; \citealt{gcn11467}). {\it Konus-Wind}
observations determined $E_{\rm peak}=490^{+103}_{-79}$ keV and
$f_{\gamma}=(3.6 \pm 0.5) \times 10^{-6}$ erg cm$^{-2}$ ($20-10^{4}$
keV; \citealt{gcn11470}). Based on the short duration and high $E_{\rm
peak}$, GRB\,101219A can be classified as a short, hard burst. The
$\gamma$-ray properties are listed in Table~\ref{tab:info}.

\subsubsection{X-ray Observations}
\label{sec:101219_xray}

\begin{figure}
\centering
\includegraphics[angle=0,width=3.2in]{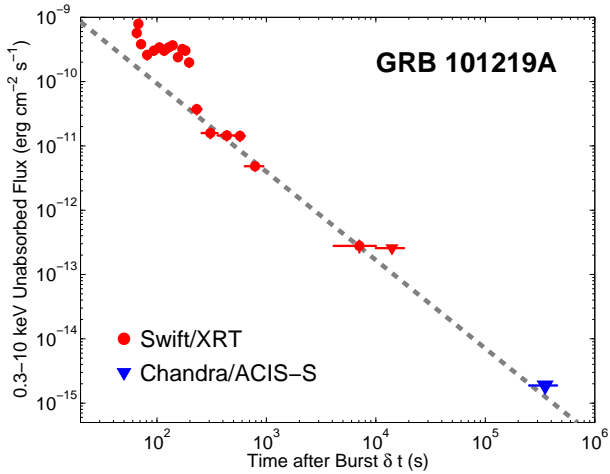}
\caption{X-ray afterglow light curve of GRB\,101219A, including
\swift/XRT observations (red points) and a {\it Chandra}/ACIS-S
observation (blue point). Triangles denote $3\sigma$ upper limits. The
data over $\delta t \approx 200-10^{4}$ s are best fit with a power
law characterized by $\alpha_X=-1.37 \pm 0.13$ (grey dashed line).
\label{fig:101219_xraylc}}
\end{figure}

\swift/XRT began observing the field at $\delta t=40$ s and detected a
fading, uncatalogued X-ray source at RA=\ra{04}{58}{20.49} and Dec=\dec{-02}{32}{23.0} with final accuracy of $1.7''$
(\citealt{gtb+07,ebp+09}; Table~\ref{tab:info}). We re-bin the XRT data and extract the best-fit
spectrum for GRB\,101219A as described in
\S\ref{sec:100625prompt}. We utilize the PC data set,
$\delta t=70-10^4$ s, where there is no evidence for spectral
evolution. We find an average best-fitting spectrum characterized by
$\Gamma=1.8 \pm 0.1$ and $N_{\rm{H,int}}=6.6^{+2.3}_{-1.8} \times
10^{21}\,\rm{cm^{-2}}$ at $z=0.718$ (C-stat$_\nu=0.97$ for $211$
d.o.f.; see \S\ref{sec:101219host} for redshift determination)
in excess of the Galactic absorption, $N_{{\rm H,MW}}=4.9\times
10^{20}\,\rm{cm^{-2}}$ \citep{kbh+05}. Our best-fit parameters are
consistent with the automatic spectrum fit produced by
\citet{gcn11474}. Applying these parameters to the XRT data, we
calculate the count rate-to-flux conversion factors, and
hence the unabsorbed fluxes (Figure~\ref{fig:101219_xraylc}).

In addition, we obtained a $20$ ks observation with the Advanced CCD
Imaging Spectrometer (ACIS-S) on-board the \chandra\ X-ray Observatory
starting at $\delta t=4.1$ days. We analyze the \chandra\ data with
the {\tt CIAO} data reduction package. In an energy range of $0.5-8$
keV, we extract $4$ counts in a $2.5''$ aperture centered on the XRT
position, consistent with the average $3\sigma$ background level
calculated from source-free regions on the same chip. We take this
count rate of $\lesssim 2 \times 10^{-4}$ counts s$^{-1}$ to be the
$3\sigma$ upper limit on the X-ray afterglow flux at $\delta t \approx
4.1$ days. Applying the spectrum extracted from the XRT data, this
count rate corresponds to $F_X \lesssim 1.9 \times 10^{-15}$ erg
cm$^{-2}$ s$^{-1}$.

The X-ray light curve is characterized by a steep decay and a short
plateau for $\delta t<200$ s, followed by a steady decline to the end
of XRT observations at $\delta t \approx 10^4$ s. To quantify this
decay rate, we utilize the single-parameter $\chi^2$-minimization
method described in \S\ref{sec:100625prompt}. Excluding the XRT
data at $\delta t \lesssim 200$ s and the late-time upper limits, the
best-fit power law index is $\alpha_X=-1.37 \pm 0.13$
($\chi^2_\nu=1.1$ for $5$ d.o.f.). The full X-ray afterglow light
curve, along with the best-fit model is shown in
Figure~\ref{fig:101219_xraylc}.

\begin{figure*}
\centering
\includegraphics[angle=0,width=6.4in]{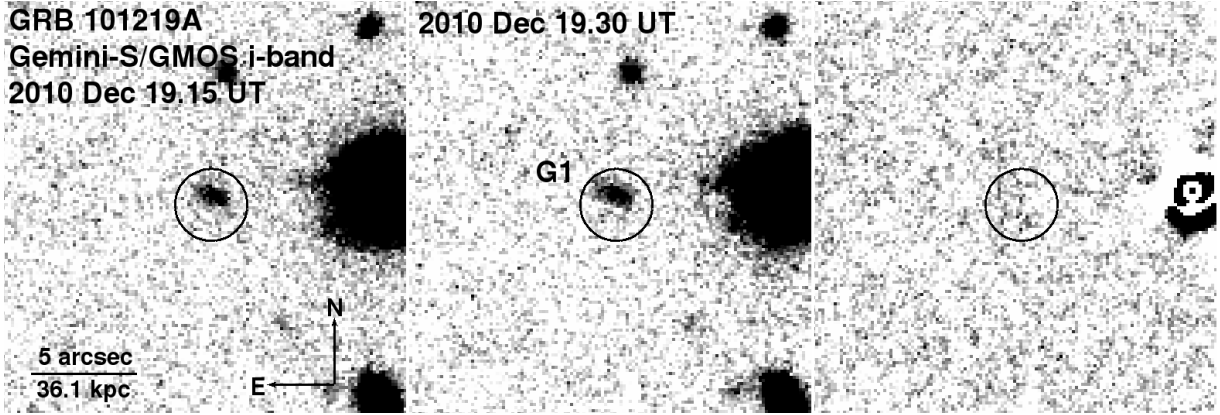}
\caption{Gemini-South/GMOS $i$-band observations of the host galaxy of
GRB\,101219A. The XRT error circle has a radius of $1.7''$ ($90\%$
containment; black). An additional $i$-band observation at $\delta
t=2.2$ hr, adds no additional constraints so is not shown here. {\it
Left:} $\delta t=0.96$ hr. {\it Center:} $\delta t=4.8$ hr. {\it
Right:} Digital image subtraction of the two epochs reveals no
afterglow to a $3\sigma$ limit of $i \gtrsim 24.9$ mag.
\label{fig:101219_i}}
\end{figure*}

\subsubsection{Optical/NIR Observations and Afterglow Limits}

UVOT commenced observations at $\delta t=67$ s. Over $\delta
t=67-5500$ s, no corresponding source was found within the XRT
position to a $3\sigma$ limit of $\gtrsim 21.4$ in the $white$ filter
\citep{gcn11472}.

We observed the field of GRB\,101219A in both $r$- and $i$-bands with
GMOS on Gemini-South, and in $J$-band with FourStar, starting at
$\delta t \approx 0.96$ hr (Table~\ref{tab:obs}). We detect a single extended
source within the XRT error circle in all filters. To assess any
fading, we obtained additional observations in
the $ri$-bands at $\delta t \approx 0.2$~d
(Table~\ref{tab:obs}). Digital image subtraction between these epochs
does not reveal any residuals (Figure~\ref{fig:101219_i}), allowing us
to place limits on the optical afterglow of $i \gtrsim 24.9$ mag and
$r \gtrsim 24.9$ mag at the time of the first epoch for each filter:
$\delta t \approx 0.96$ and $2.2$ hr, respectively
(Table~\ref{tab:obs}). To assess the fading on timescales $\gtrsim 1$
day, we obtained a third set of observations in the $r$-band at
$\delta t \approx 9$~d. Image subtraction with each of the first and
second $r$-band observations also show no evidence for fading
(Table~\ref{tab:obs}). A second set of $J$-band observations
at $\delta t \approx 350$~d and a clean image subtraction with the
first epoch allows us to place a limit on the NIR afterglow of $J
\gtrsim 23.6$ mag at $\delta t=1.7$ hr. Finally, to complement our early
optical/NIR observations, we obtained imaging of the putative host
galaxy in the $gz$-bands with LDSS3 starting on 2011 January 12.15~UT,
and in the $K_s$-band with FourStar on 2011 December 07.24~UT. Our
limits for the afterglow and photometry of the putative host galaxy
are summarized in Table~\ref{tab:obs}.

We obtained spectroscopic observations of the host on 2011
January 2.25~UT using GMOS on Gemini-North at a mean airmass of
$1.2$. We obtained a set of $4 \times 1800$ s exposures with the
R$400$ grating and an order-blocking filter, OG515 in the
nod-and-shuffle mode, covering $5860-10200$ \AA\ at a spectral
resolution of $\approx 7$ \AA. We used standard tasks in IRAF for data
reduction, CuAr arc lamps for wavelength calibration, and archival
observations of the smooth-spectrum standard star BD+28 4211 for flux
calibration. We discuss the characteristics of the spectrum and
redshift determination in \S\ref{sec:101219host}.

\subsection{GRB\,110112A}

\begin{figure}
\centering
\includegraphics[angle=0,width=3.2in]{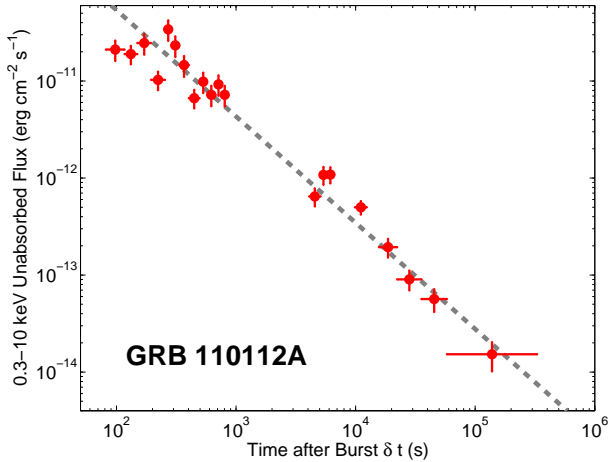}
\caption{\swift/XRT light curve of GRB\,110112A. The data (red
points) for $\delta t \gtrsim 200$ s is best fit with a single power
law characterized by $\alpha_X=-1.10 \pm 0.05$ (grey dashed line).
\label{fig:110112a_xraylc}}
\end{figure}

{\it Swift}/BAT detected GRB\,110112A on 2011 January 12.175 UT
\citep{gcn11553}, with a single spike with
$T_{90} = 0.5 \pm 0.1$ s ($15-350$ keV) and $f_{\gamma} = (3.0 \pm 0.9)
\times 10^{-8}$ erg cm$^{-2}$ ($15-150$ keV; \citealt{gcn11557}). The
BAT ground-calculated position is RA=\ra{21}{59}{33.6} and
Dec=\dec{+26}{28}{10.6} with $2.6'$ radius uncertainty ($90\%$
containment; \citealt{gcn11557}). The $\gamma$-ray properties are
listed in Table~\ref{tab:info}.

\subsubsection{X-ray Observations}

XRT commenced observations of the field of GRB\,110112A at $\delta
t=76$ s and located a fading X-ray counterpart with a UVOT-enhanced
positional accuracy of $1.6''$ radius
(\citealt{gcn11556,gtb+07,ebp+09}; Table~\ref{tab:info}). We extract the XRT light curve and
spectrum in the manner described in Section~\ref{sec:100625prompt},
requiring a minimum $S/N=3$ for each bin, and use the Galactic
absorption in the direction of the burst of $N_{\rm H,MW} = 5.5 \times
10^{20}$ cm$^{-2}$ \citep{kbh+05}. The light curve is characterized by
a short plateau for $\delta t \lesssim 200$~s, followed by a steady
decline (Figure~\ref{fig:110112a_xraylc}). Performing
$\chi^2$-minimization, we find the XRT light curve for $\delta t
\gtrsim 200$~s is best fit with a single power law characterized by
index $\alpha_X=-1.10 \pm 0.05$ ($\chi^2_{\nu}=1.0$ for $17$
d.o.f.). Our best-fitting spectral parameters over the entire data
set, where there is no evidence for spectral evolution, are $\Gamma =
2.2 \pm 0.2$ and an upper limit of $N_{\rm H,int} \lesssim 1.6 \times
10^{21}$ cm$^{-2}$ ($3\sigma$ at $z=0$; C-stat $=0.82$ for $156$
d.o.f.).

\subsubsection{Optical Afterglow Discovery}

\begin{figure}
\centering
\includegraphics*[angle=0,width=3.4in]{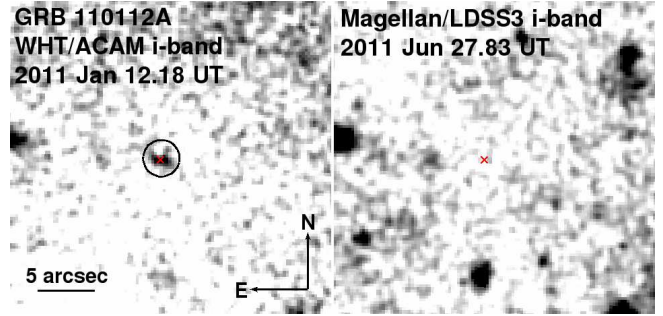}
\caption{The optical afterglow of GRB\,110112A. The XRT error circle
has a radius of $1.6''$ ($90\%$ containment; black) and the red cross
marks the centroid of the optical afterglow, with a $1\sigma$
uncertainty of $0.14''$ (afterglow centroid + absolute tie to SDSS)
and $i=22.77 \pm 0.29$. {\it Left:} WHT/ACAM $i$-band observations at
$\delta t=0.64$ days. {\it Right:} Magellan/LDSS3 $i$-band
observations at $\delta t=166$ days.
\label{fig:110112A_sub}}
\end{figure}

UVOT commenced observations at $\delta t=80$ s, and no corresponding
source was found within the XRT position to a $3\sigma$ limit in the
$white$ filter of $\gtrsim 21.3$ mag using data over $\delta t=4400-6100$~s
(uncorrected for Galactic extinction; \citealt{gcn11560}).

We obtained $i$-band observations with ACAM mounted on the $4.2$-m
William Herschel Telescope (WHT) at $\delta t=15.4$ hr. In a total
exposure time of $600$~s (Table~\ref{tab:obs}), we detect a single
source within the enhanced XRT error circle with $i=22.77 \pm 0.29$
mag, where the zeropoint has been determined using sources in common
with the SDSS catalog (Figure~\ref{fig:110112A_sub}). To assess any
fading associated with this source or within the XRT position, we
obtained $i$-band imaging with LDSS3 starting on 2011 June 27.83~UT and
no longer detect any source within the error circle to $i \gtrsim
24.7$ mag, confirming that the source has faded by $\gtrsim 2$
mag. Therefore, we consider this source to be the optical afterglow of
GRB\,110112A.

To determine the position of the afterglow, we perform absolute
astrometry using $108$ point sources in common with SDSS and calculate
an astrometric tie RMS of $0.11''$. The resulting afterglow position
is RA=\ra{21}{59}{43.85} and Dec=\dec{+26}{27}{23.89} (J2000) with a
centroid uncertainty of $0.09''$ determined with Source Extractor,
which, together with the astrometric tie uncertainty, gives a total
positional uncertainty of $0.14''$. We note that this source's
position is not consistent with the $R = 19.6 \pm 0.3$ source claimed
by \citet{gcn11554}. Furthermore, we do not detect any source at this
position in any of our observations.

To perform a more thorough search for a coincident host galaxy, we
obtained $r$-band observations with LDSS3 on 2011 June 27.83~UT and
$i$-band observations with Gemini-North/GMOS on 2011 July 28.46~UT. In
these deeper observations, we do not detect any sources within the XRT
error circle to limits of $r \gtrsim 25.5$ mag and $i \gtrsim 26.2$
mag (Table~\ref{tab:obs}). We further assess the probability of
potential host galaxies outside the XRT position in
\S\ref{sec:110112a_cc}.

\section{Afterglow Properties}

We utilize the X-ray and optical/NIR observations to constrain the
explosion properties and circumburst environments of GRBs\,100625A,
101219A and 110112A.  We adopt the standard synchrotron model for a
relativistic blastwave in a constant density medium (ISM), as expected
for a non-massive star progenitor \citep{sph99,gs02}. This model
provides a mapping from the broad-band afterglow flux densities to
physical parameters: isotropic-equivalent kinetic energy ($E_{\rm
K,iso}$), circumburst density ($n_0$), fractions of post-shock energy
in radiating electrons ($\epsilon_e$) and magnetic fields
($\epsilon_B$), and the electron power-law distribution index, $p$,
with $N(\gamma)\propto \gamma^{-p}$ for $\gamma \gtrsim \gamma_{\rm
min}$. Since we have optical and X-ray observations for these three
bursts, we focus on constraining the location of the cooling frequency
($\nu_c$) with respect to the X-ray band because it affects the
afterglow flux dependence on $E_{\rm K,iso}$ and $n_0$. For each
burst, we determine this by comparing the temporal ($\alpha_X$) and
spectral ($\beta_X \equiv 1-\Gamma$) indices to the closure relation
$\alpha-3\beta/2$: for $p>2$, if $\nu_c>\nu_X$, $\alpha-3\beta/2=0$,
while for $\nu_c<\nu_X$, $\alpha-3\beta/2=1/2$. We also infer the
extinction, $A_V^{\rm host}$ by a comparison of the optical and X-ray
data.

\subsection{GRB\,100625A}

From the X-ray light curve and spectrum of GRB\,100625A, we measure a
temporal decay index of $\alpha_X=-1.45 \pm 0.08$ and a spectral index of
$\beta_X=-1.5 \pm 0.2$, which gives $\alpha_X-3\beta_X/2=0.79 \pm
0.34$. This indicates that $\nu_c<\nu_X$ and therefore $p=2.7 \pm
0.2$.

From our derived value of $N_{\rm H,int} \lesssim 1.7 \times 10^{21}$
cm$^{-2}$, we infer $A_{V}^{\rm host}\lesssim 0.8$ mag ($3\sigma$) in
the rest-frame of the burst using the Galactic $N_H$-to-$A_V$
conversion, $N_{\rm H,int}/A_V\approx 2.0 \times 10^{21}$
\citep{ps95,wat11}. We can also investigate the presence of extinction
by comparing the X-ray flux and the optical upper limit at $\delta t
\approx 0.5$~d. If we assume a maximum value of $\nu_{c,{\rm max}}
\approx 2.4 \times 10^{17}$~Hz ($1$ keV) and extrapolate the X-ray
flux density of $\approx 9 \times 10^{-3}\,\mu$Jy to the optical band using
$\beta=-(p-1)/2=-0.85$ to obtain the lowest bound on the expected
afterglow flux in the absence of extinction, we estimate $F_{\nu,{\rm
opt}} \approx 0.24\,\mu$Jy ($i=25.4$ mag). Given the observed
limit of $F_{\nu,{\rm opt}} \lesssim 2.9\,\mu$Jy ($i \gtrsim 22.7$
mag), this does not conflict with this lower bound, the afterglow
observations are consistent with no extinction.

We can therefore use the X-ray data and optical afterglow limits to
constrain $E_{\rm K,iso}$ and $n_0$. Assuming that the X-ray flux is
from the forward shock, we can directly obtain $E_{\rm K,iso}$ by \citep{gs02}

\begin{equation}
E_{\rm K,iso,52}^{4.7/4} \epsilon_{e,-1}^{1.7} \epsilon_{B,-1}^{0.7/4} \approx 5.7 \times 10^{-3},
\end{equation}

\noindent where $E_{\rm K,iso,52}$ is in units of $10^{52}$ erg, and
$\epsilon_e$ and $\epsilon_B$ are in units of $10^{-1}$, and we have used
$z=0.452$. The X-ray flux density at $\delta t \approx 10^{4}$~s is $F_{\nu,X}
\approx 9.1 \times 10^{-3}\,\mu$Jy ($1$ keV), and therefore $E_{\rm
K,iso} \approx 1.2 \times 10^{50}$ erg
($\epsilon_e=\epsilon_B=0.1$). At $z=0.452$, $E_{\rm \gamma,iso}
\approx 4.3 \times 10^{50}$ erg ($20-2000$ keV from the {\it
Konus-Wind} fluence), which gives a $\gamma$-ray efficiency of
$\eta_\gamma \approx 0.8$. If we instead assume $\epsilon_e=0.1$ and
$\epsilon_B=0.01$, then $E_{\rm K,iso} \approx 1.7 \times 10^{50}$
erg, and $\eta_\gamma \approx 0.7$.

For $\nu_m<\nu_{\rm opt}<\nu_c$ (where $\nu_m$ is the synchrotron peak frequency),
the optical afterglow brightness depends on a combination of $E_{\rm
K,iso}$ and $n_0$. Therefore, the $riz$-band limits on the afterglow
translate to an upper limit on the physical parameters, given by

\begin{equation}
E_{\rm K,iso,52}^{5.7/4} n_0^{0.5} \epsilon_{e,-1}^{1.7} \epsilon_{B,-1}^{3.7/4} \lesssim 2.5 \times 10^{-3},
\end{equation}

\noindent where $n_0$ is in units of cm$^{-3}$. Assuming
$\epsilon_e=\epsilon_B=0.1$ and using $E_{\rm K,iso}=1.2 \times
10^{50}$ erg, we obtain $n \lesssim 1.5$ cm$^{-3}$. If we instead
assume $\epsilon_e=0.1$ and $\epsilon_B=0.01$, then $n_0 \lesssim 40$
cm$^{-3}$. For both scenarios, we obtain $\nu_c \gtrsim 4 \times
10^{15}$ Hz ($\gtrsim 0.02$ keV), consistent with our assumption that
$\nu_c<\nu_X$.

\subsection{GRB\,101219A}

From the X-ray light curve and spectrum, we measure $\alpha_{X}=-1.37
\pm 0.13$ and $\beta_{X}=-0.8 \pm 0.1$, which gives
$\alpha_X-3\beta_X/2=0.17 \pm 0.23$, suggesting that
$\nu_c>\nu_X$. The resulting value of $p$ is $2.7 \pm 0.1$. We note
that the closure relation is consistent with the alternative scenario
for $>2\sigma$.

Since the optical afterglow flux may be subject to an appreciable
amount of extinction, as suggested by the intrinsic absorption in the
X-ray spectrum (\S\ref{sec:101219_xray}), the most reliable proxy
for $E_{\rm K,iso}$ and $n_0$ is the X-ray afterglow flux. Using the
last XRT data point at $\delta t \approx 7 \times 10^{3}$~s, which has
$F_{\nu,X}\approx 0.03\,\mu$Jy ($1$ keV), we infer the following
relationship between $E_{\rm K,iso}$ and $n_0$,

\begin{equation}
E_{\rm K,iso,52}^{5.7/4} n_0^{0.5} \epsilon_{e,-1}^{1.7} \epsilon_{B,-1}^{3.7/4} \approx 1.3 \times 10^{-3},
\end{equation}
\label{eqn:fom}

\noindent where we have used $z=0.718$. At this redshift, we find
$E_{\rm \gamma,iso} \approx 4.8 \times 10^{51}$ erg ($20-10^4$ keV
using the {\it Konus-Wind} fluence). Assuming $E_{\rm \gamma,iso}
\approx E_{\rm K,iso}$, we infer $n_0 \approx 1.3 \times 10^{-5}$
cm$^{-3}$ for $\epsilon_e=\epsilon_B=0.1$. With these values, $\nu_c
\approx 6 \times 10^{19}$ Hz ($250$ keV), consistent with our
assumption that $\nu_c>\nu_X$. We note that this assumption is
violated for $n_0 \gtrsim 4 \times 10^{-3}$ cm$^{-3}$. If instead we
use $\epsilon_e=0.1$ and $\epsilon_B=0.01$, then we obtain $n_0
\approx 9 \times 10^{-4}$ cm$^{-3}$ and $\nu_c \approx 2 \times
10^{19}$ Hz ($80$ keV), which is again self-consistent, and find this
assumption is violated for $n_0 \gtrsim 0.1$ cm$^{-3}$. Therefore, the
X-ray data suggest an explosion environment with $n_0 \approx
10^{-5}-10^{-3}$ cm$^{-3}$ for GRB\,101219A.

We investigate the presence of extinction intrinsic to the host galaxy
by comparing the X-ray and NIR observations, since the NIR data
provide a stronger constraint than the optical band. Since the X-ray
and NIR bands lie on the same segment of the synchrotron spectrum, the
spectral slope is given by $\beta_{\rm NIR-X}=\beta_X \approx
-0.8$. At the time of our first $J$-band observations at $\delta t
\approx 1$ hr, the X-ray flux density is $0.06\,\mu$Jy, leading to an
expected $J$-band flux density of $F_{\nu,J} \approx 14.7\,\mu$Jy ($21$
mag). This is above the limit of our observations, $\lesssim
1.4\,\mu$Jy ($\gtrsim 23.6$ mag), indicating that $A_J \gtrsim 2.5$
mag. Using a Milky Way extinction curve \citep{ccm89}, this indicates
that $A_V^{\rm host} \gtrsim 4.2$ mag in the rest-frame of the
burst. In addition, using the Galactic relation between $N_H$ and
$A_V$, this implies $N_{\rm H,int} \gtrsim 7.5 \times 10^{21}$
cm$^{-2}$, which does not necessarily violate our inferred value from
the X-ray spectrum of $N_{\rm H,int}=(6.6 \pm 2.0) \times 10^{21}$
cm$^{-2}$. Therefore, the broad-band afterglow spectrum requires an
appreciable amount of extinction.

\subsection{GRB\,110112A}

From the X-ray light curve and spectrum, we measure $\alpha_{X}=-1.10
\pm 0.05$ and $\beta_{X}=-1.2 \pm 0.2$, giving
$\alpha_X-3/2\beta_X=0.70 \pm 0.30$ indicating $\nu_c<\nu_X$. The
resulting value of $p$ is $2.1 \pm 0.1$.

From our derived value of $N_{\rm H,int} \lesssim 1.6 \times 10^{21}$
cm$^{-2}$, we infer $A_{V}^{\rm host}\lesssim 0.9$ mag in the
rest-frame of the burst using the Galactic relation. We can
measure the cooling frequency by comparing the X-ray and
optical fluxes at $\delta t \approx 0.64$~d. At this time, $F_{\nu,X}
\approx 6.6 \times 10^{-3}\,\mu$Jy and $F_{\nu,{\rm opt}} \approx
2.8\,\mu$Jy. Using $p=2.1$ and the location of the optical and X-ray
bands, we then estimate that $\nu_c \approx 1.6 \times 10^{15}$ Hz
($\approx 7 \times 10^{-3}$ keV) which agrees with our assumption that
$\nu_c<\nu_X$. The cooling frequency is dependent on a combination of
physical parameters and gives the constraint:

\begin{equation}
E_{\rm K,iso,52}^{-0.5} n_0^{-1} \epsilon_{B,-1}^{-1.5} \approx 5.4,
\end{equation}

\noindent where we have assumed a fiducial redshift of $z=0.5$, the
median of the observed short GRB population. We then use the X-ray
afterglow flux at $\delta t \approx 0.64$~d to determine $E_{\rm
K,iso}$ by

\begin{equation}
E_{\rm K,iso,52}^{4.1/4} \epsilon_{e,-1}^{1.1} \epsilon_{B,-1}^{0.1/4} \approx 0.023.
\end{equation}

\noindent Our final constraint comes from the optical afterglow brightness, given by

\begin{equation}
E_{\rm K,iso,52}^{5.1/4} n_0^{0.5} \epsilon_{e,-1}^{1.1} \epsilon_{B,-1}^{3.1/4} \approx 0.01.
\end{equation}
\label{eqn:fom}

\noindent Assuming $\epsilon_e=0.1$ and $z=0.5$, we obtain the
solution $E_{\rm K,iso} \approx 2.5 \times 10^{50}$ erg, $n_0 \approx
1.5$ cm$^{-3}$ and $\epsilon_B \approx 0.08$. At this redshift,
$E_{\rm \gamma,iso} \approx 9.5 \times 10^{49}$ erg (determined from
the \swift\ fluence and applying a correction factor of $5$ to
represent $\approx 1-10^4$ keV). If we consider a high-redshift origin
for GRB\,110112A of $z=2$, then we infer larger energies of $E_{\rm
K,iso} \approx 3.6 \times 10^{51}$ erg and $E_{\rm \gamma,iso} \approx
1.5 \times 10^{51}$ erg, a lower value of $\epsilon_B \approx 0.01$,
and a lower density, $n_0 \approx 0.18$ cm$^{-3}$. In both cases,
$\eta_\gamma \approx 0.3$.

\section{Host Galaxy Properties}

\subsection{GRB\,100625A}
\label{sec:100625cc}
\label{sec:100625spec}
\label{sec:100625sed}

The XRT position of GRB\,100625A fully encompasses a single galaxy,
which we call G1 (Figure~\ref{fig:100625}). To assess the probability
that the burst originated from G1, we calculate the probability of
chance coincidence, $P_{cc}(<\delta R)$, at a given angular
separation, ($\delta R$) and apparent magnitude ($m$) for galaxies
within $15'$ (the field of view of our images) of the burst position
\citep{bkd02,ber10}. For G1, we conservatively assume $\delta R =
3\sigma_{\rm XRT} \approx 3.4''$, and calculate $P_{cc}(<\delta
R)\approx 0.04$. The remaining bright galaxies in the field have
substantially higher values of $P_{cc}(<\delta R) \gtrsim 0.17$, and a
search for galaxies within $5^{\circ}$ of the GRB position
using the NASA/IPAC Extragalactic Database (NED) yields only objects
with $P_{cc}\gtrsim 0.98$. From these probabilistic arguments, we
consider G1 to be the host galaxy of GRB\,100625A.

\begin{figure}
\centering
\includegraphics*[angle=0,width=3.2in]{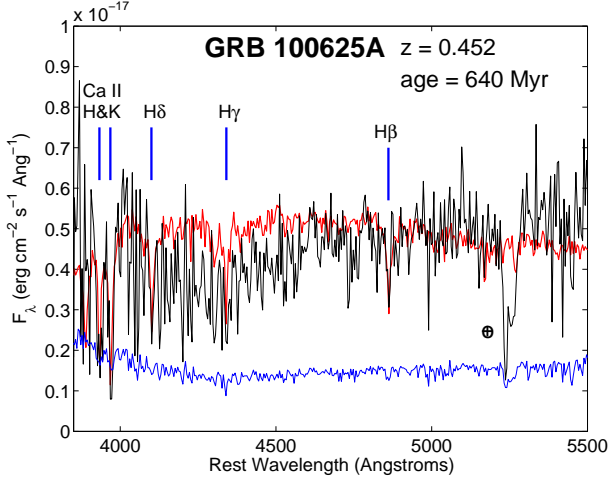}
\caption{LDSS3 spectrum of the early-type host galaxy of GRB\,100625A, binned
with a 3-pixel boxcar (black: data; blue: error spectrum). Also shown
is the best-fit SSP template (red; \citealt{bc03}) with a stellar
population age of $640$ Myr at a redshift of $z=0.452 \pm 0.002$.  Fits are
performed on the unbinned data. The locations of the Balmer absorption
lines and \ion{Ca}{2} H\&K are labelled. 
\label{fig:100625spec}}
\end{figure}

To determine the host galaxy's redshift, we fit the LDSS3 spectrum
over the wavelength range of $5200-8000$ \AA\ with simple stellar
population (SSP) spectral evolution models at fixed ages ($\tau=0.29,
0.64, 0.90, 1.4$ and $2.5$ Gyr) provided as part of the GALAXEV
library \citep{bc03}; at wavelengths outside this range, the
signal-to-noise is too low to contribute significantly to the fit. We
use $\chi^{2}$-minimization with redshift as the single free
parameter, and perform the fit on the unbinned data. The resulting
best-fit redshift is $z=0.452 \pm 0.002$ ($\chi^{2}_{\nu}=1.3$ for
$1861$ degrees of freedom), determined primarily by the location of
the $4000$\,\AA\ break and the main absorption features of \ion{Ca}{2}
H\&K, H$\beta$, H$\gamma$, and H$\delta$. The shape of the break is
best fit by the template with $\tau=0.64$~Gyr
(Figure~\ref{fig:100625spec}), and poorer fits ($\chi^{2}_{\nu}
\gtrsim 2$) are found for SSPs with younger or older ages. Due to the
strength of the $4000$ \AA\ break, deep absorption features, lack of emission
lines, and old age, we classify this host as an early-type galaxy.

We do not find an emission feature corresponding to
[\ion{O}{2}]$\lambda 3727$. Using the error spectrum, we calculate the
expected integrated flux for a $3\sigma$ emission doublet centered at
$\lambda=3727$\,\AA\ with a width of $\approx 10$\,\AA. We find an
expected upper limit of $F_{\rm [OII]} \lesssim 4.3\times 10^{-17}$ erg
cm$^{-2}$ s$^{-1}$, which translates to $L_{\rm [OII]} \lesssim
2.2\times 10^{40}$ erg s$^{-1}$ at the redshift of the burst. Using
the standard relation, ${\rm SFR}=(1.4 \pm 0.4)$ M$_\odot$
yr$^{-1}$\,L$_{\rm [OII],41}$ \citep{ken98}, we derive a $3\sigma$
upper limit of SFR$\lesssim 0.3\,M_{\odot}$ yr$^{-1}$ for the host
galaxy.

\begin{figure}
\centering
\includegraphics*[angle=0,width=3.2in]{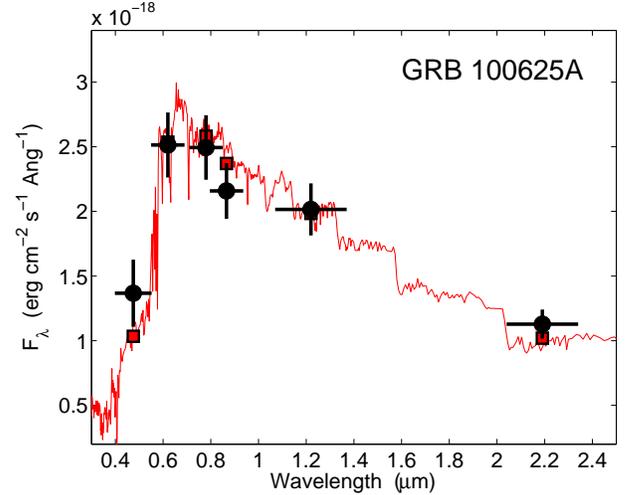}
\caption{$grizJK_s$ photometry for the host galaxy of GRB\,100625A
(black circles). The best-fit model (red squares and line;
\citealt{mar05}) is characterized by $\tau \approx 0.8$ Gyr and $M_*
\approx 4.6 \times 10^{9}\,M_{\odot}$.
\label{fig:100625sed}}
\end{figure}

We use the $grizJK_s$-band photometry to infer the stellar population
age and mass of the host galaxy with the \citet{mar05} evolutionary
stellar population synthesis models, employing a Salpeter initial mass
function and a red giant branch morphology. We fix $A_{V}^{\rm
host}=0$ mag as inferred from the absence of $N_{\rm H,int}$
(\S\ref{sec:100625prompt}), $z=0.452$ as inferred from the spectrum,
and metallicity $Z=Z_{\odot}$, and allow the stellar population age
($\tau$) and stellar mass ($M_*$) to vary. The resulting best-fit
model is characterized by $\tau \approx 0.8$ Gyr, in good agreement
with the fit to the spectrum, and $M_* \approx 4.6\times
10^{9}\,M_{\odot}$. The model and broad-band photometry are shown in
Figure~\ref{fig:100625sed}.

\subsection{GRB\,101219A}
\label{sec:101219host}

The XRT position of GRB\,101219A fully encompasses a single galaxy
(G1; Figure~\ref{fig:101219_i}). We perform the same probability of
chance coincidence analysis described in \S\ref{sec:100625cc}
using $\delta R = 3\sigma_{\rm XRT}$ and find $P_{cc}(<\delta
R)\approx 0.06$ for G1, while the remaining bright galaxies within
$5'$ of the burst have $P_{cc}(<\delta R)\gtrsim 0.23$. Furthermore, a
search within $5^{\circ}$ of the position with NED yields only
galaxies with $P_{cc}(<\delta R) \approx 1$. We therefore consider G1
to be the most probable host galaxy of GRB\,101219A.

\begin{figure}
\centering
\includegraphics*[angle=0,width=3.2in]{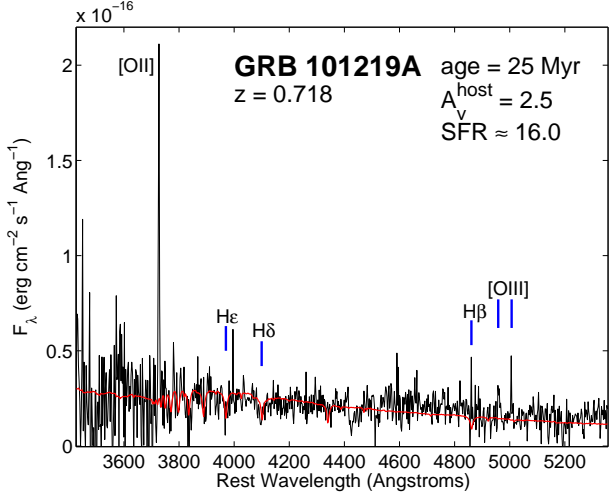}
\caption{GMOS-N spectrum of the host galaxy of GRB\,101219A, binned
with a 3-pixel boxcar (black). The spectrum is corrected for Galactic
extinction and $A_V^{\rm host}=2.5$ mag. The stellar population model
has $\tau=25$ Myr (red; \citealt{bc03}). The [\ion{O}{2}] $\lambda
3727$ and [\ion{O}{3}] $\lambda 5007$ emission features are at a
common redshift of $z=0.718$. Also labeled are the locations of the
Balmer lines H$\epsilon$ and H$\delta$, and marginal emission features
at H$\beta$ and the [\ion{O}{3}] doublet. From [\ion{O}{2}] $\lambda
3727$ we deduce SFR $=16.0 \pm 4.6$ M$_{\odot}$ yr$^{-1}$.
\citep{ken98}.
\label{fig:101219spec}}
\end{figure}

We examine the host spectrum of GRB\,101219A to determine the redshift
and physical characteristics of the stellar population. We identify
two emission features in the co-added spectrum at $\lambda_{\rm
obs}=6401.65$\,\AA\ and $\lambda_{\rm obs}=8599.50$\,\AA\ that are also
present in the individual 2D spectra prior to co-addition. If these
features correspond to [\ion{O}{2}]$\lambda 3727$ and
[\ion{O}{3}]$\lambda 5007$, their locations give a
common redshift of $z=0.718$. Furthermore, we do not find a common redshift
solution for an alternative set of features, so we consider the
host galaxy to be at $z=0.718$. In addition, we note the presence of
marginal emission features at the expected locations of H$\beta$ and
[\ion{O}{3}]$\lambda 4959$; however, these locations are contaminated
by sky line residuals. Finally, we detect absorption at the locations
of $H\varepsilon$ and $H\delta$ (Figure~\ref{fig:101219spec}).

To determine the age and host extinction, we use stellar population
spectral templates with fixed ages of $\tau=5, 25, 100$ and $290$ Myr
\citep{bc03} to fit the continuum; ages outside this range do not fit
the overall shape of the spectrum. We apply corrections for both Galactic
extinction ($A_V=0.16$ mag at $z=0$; \citealt{sf11}) and $A_V^{\rm
host}$ at $z=0.718$ using a Milky Way extinction curve
\citep{ccm89}. The spectrum is best matched with the $\tau=25$ Myr
template and $A_V^{\rm host}=2.5$ mag. Since there is some degeneracy
between age and $A_V^{\rm host}$, imposing an older stellar population
of $\tau=100$ Myr also provides a reasonable match, but requires a
smaller amount of extinction of $A_V^{\rm host} \approx 2$ mag. Older
spectral templates predict a large break at $4000$ \AA\ not seen in the
spectrum, while younger templates lack the observed absorption
lines. Therefore, a likely range of ages for the host galaxy is $\tau
\approx 25-100$ Myr. Given the emission features and relatively young
age, we classfiy this galaxy as late-type. The de-reddened spectrum
for GRB\,101219A, along with the $25$~Myr model, is shown in
Figure~\ref{fig:101219spec}.

From the extinction-corrected flux of [\ion{O}{2}]$\lambda 3727$,
$F_{\rm [OII]}\approx 8.5\times 10^{-16}$ erg cm$^{-2}$ s$^{-1}$, we find
$L_{\rm [OII]}\approx 1.1 \times 10^{42}$ erg s$^{-1}$ at the redshift
of the burst. Using the standard relation \citep{ken98}, we derive a
SFR of $16.0 \pm 4.6$ M$_\odot$ {\rm yr}$^{-1}$.

\begin{figure}
\centering
\includegraphics*[angle=0,width=3.2in]{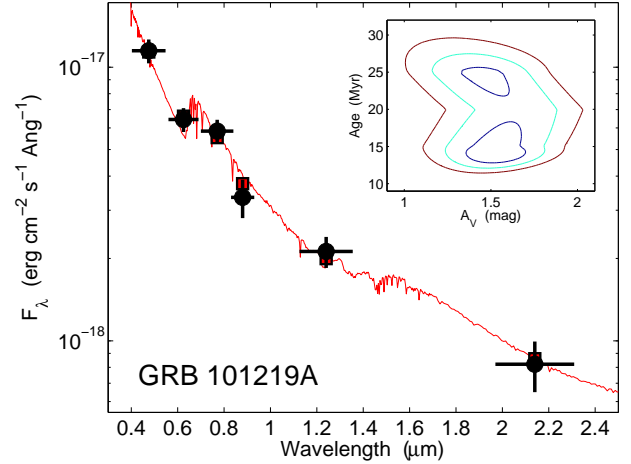}
\caption{$grizJK_s$-band photometry of the host galaxy of GRB\,101219A
(black circles). The best-fit model (red squares and line;
\citealt{mar05}) is characterized by $A_V^{\rm host} \approx 1.5$ mag,
$\tau \approx 15-25$ Myr and $M_* \approx 1.4 \times
10^{9}\,M_{\odot}$. The age-$A_V^{\rm host}$ contours of $1\sigma$ (blue),
$2\sigma$ (cyan), and $3\sigma$ (red) solutions are shown in the inset.
\label{fig:101219sed}}
\end{figure}

We use the same procedure described in \S\ref{sec:100625sed} to
model the SED of the host galaxy to infer $\tau$ and $M_*$. We fix
$z=0.718$ as inferred from the spectrum, $Z=Z_{\odot}$, and allow
$\tau$, $M_*$, and $A_{V}^{\rm host}$ to vary. The resulting best-fit
model is characterized by $A_V^{\rm host} \approx 1.5$ mag, $\tau
\approx 15-25$ Myr, and $M_* \approx 1.4 \times 10^{9}\,M_{\odot}$,
which is consistent with the parameters derived from the spectrum and
afterglow. The broad-band photometry and best-fit stellar population
model are shown in Figure~\ref{fig:101219sed}.

\subsection{GRB\,110112A}
\label{sec:110112a_cc}

\begin{figure}
\centering
\includegraphics*[angle=0,width=3.2in]{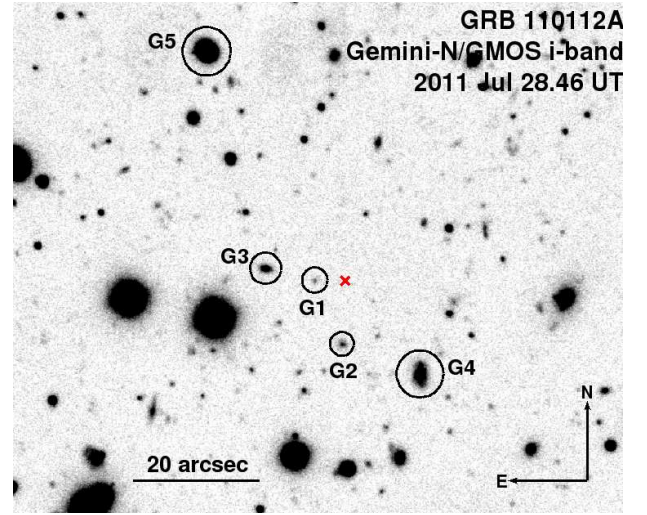}
\caption{{\it Left:} Gemini-N/GMOS $i$-band observations of the field
of GRB\,110112A on 2011 Jul 28.46~UT. The position of the optical
afterglow is marked by the red cross. The five galaxies with the lowest
probabilities of chance coincidence are circled and labeled G1-G5. The galaxy
with the lowest value of $P(<\delta R)$ is G1, located $4.8''$ from the
optical afterglow position.
\label{fig:110112a_field}}
\end{figure}

\begin{figure}
\centering
\includegraphics*[angle=0,width=3.2in]{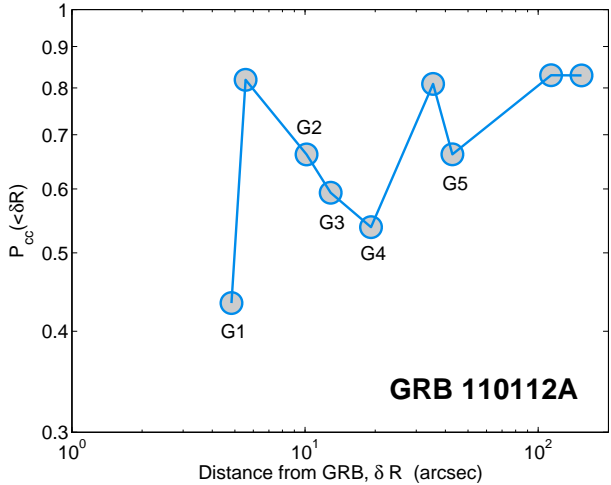}
\caption{Probability of chance coincidence, $P_{cc}(<\delta R)$, as a
function of angular distance from the optical afterglow position of
GRB\,110112A. There are nine galaxies in the $2'$ field with
$P_{cc}(<\delta R)<0.85$. The five galaxies with the lowest $P(<\delta
R)$ are labeled G1-G5. The galaxy G1 has the lowest probability of
chance coincidence $P(<\delta R)=0.43$.
\label{fig:110112a_cc}}
\end{figure}

For GRB\,110112A, we do not detect a source in coincidence with the
optical afterglow position or within the XRT error circle to a
$3\sigma$ limit of $i \gtrsim 26.2$ mag in our GMOS-N image
(Figure~\ref{fig:110112a_field}). To determine which sources in the
field are probable hosts, we calculate $P_{cc}(<\delta R)$ for $15$
galaxies within $\sim 3'$ of the GRB position, the field
of view of our GMOS-N $i$-band image. These galaxies were selected by
discarding noticeably fainter galaxies with increasing $\delta R$
since these objects will have $P_{cc}(<\delta R) \sim 1$. We find that
$9$ of these galaxies have $P_{cc}(<\delta R) \lesssim 0.85$
(Figure~\ref{fig:110112a_cc}). The two most probable host galaxies, G1
and G4 (Figures~\ref{fig:110112a_field} and ~\ref{fig:110112a_cc}),
have $P_{cc}(<\delta R)=0.43$ and $0.54$, respectively, and offsets of
$\delta R=4.8''$ and $11.1''$.  In addition, we search for bright
galaxies within $5^{\circ}$ of the GRB position using NED, but all
additional catalogued galaxies have $P_{cc}(<\delta R)\gtrsim
0.98$. Given the relatively high values for $P_{cc}(<\delta R)$, we do
not find a convincing putative host for GRB\,110112A.

It is also plausible that GRB\,110112A originated from a galaxy
fainter than the detection threshold of our observations. For
instance, a $\approx 27$ mag host would require $\delta R \lesssim
2.0''$ while a $\approx 28$ mag host would require $\delta R \lesssim
1.3''$, to be a more probable host than G1. However, to be a $27-28$
mag galaxy convincing enough to make a host association
($P_{cc}(<\delta R) \lesssim 0.05$) would require a smaller offset of
$\delta R \lesssim 0.5''$. We note that the lack of potential host is
in contrast to previous ``host-less'' short GRBs \citep{ber10}. The
high inferred density due to the bright optical afterglow (\S3) is
suggestive of a high-redshift origin as opposed to a progenitor system that
was kicked outside of its host galaxy.

\section{Stellar Population Characteristics}

\begin{figure}
\centering
\includegraphics[angle=0,width=3.2in]{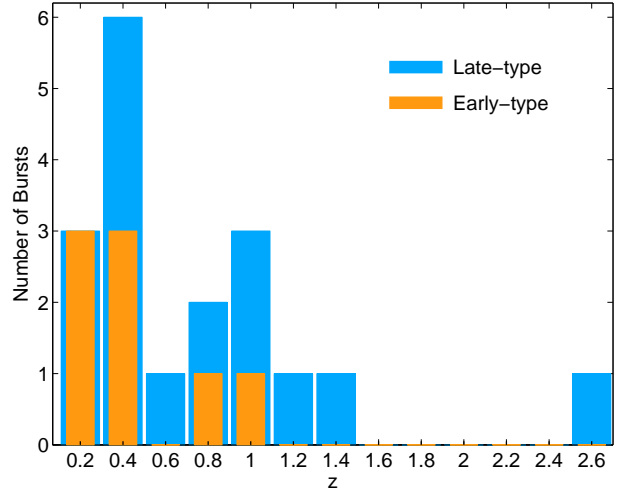}
\caption{Redshift distribution of $26$ short GRBs with host
associations and redshifts, classified by type of the host galaxy,
either late-type (blue) or early-type (orange).
\label{fig:zhist}}
\end{figure}

Of the $30$ short GRBs with host associations ($P_{cc}\lesssim 0.05$;
Table~\ref{tab:dem}), GRB\,100625A is the fifth short GRB associated
with a spectroscopically-confirmed early-type host galaxy
\citep{gso+05,bpc+05,bpp+06,bpc+07,fbc+11}, near the median redshift
of the short GRB population (Figure~\ref{fig:zhist}). In contrast,
GRB\,101219A is associated with a $z=0.718$ late-type galaxy that is
actively star-forming with characteristics similar to the majority of
the short GRB late-type host population \citep{ber09}. Finally,
GRB\,110112A joins a growing number of short GRBs with sub-arcsecond
positions but no obvious coincident host galaxy to deep limits of
$\gtrsim 26$ mag \citep{ber10}, although unlike previous events, the
case for a large offset is less clear.

Short GRBs with sub-arcsecond positions and coincident hosts have a
median projected physical offset of $\sim 5$ kpc \citep{fbf10} which,
in the context of a NS-NS/NS-BH progenitor, can be interpreted as the
result of natal kicks and moderate delay times
\citep{fwh99,bpb+06}. At the inferred redshifts of GRBs\,100625A and
101219A, the upper limits on the projected physical offsets set by the
radii of the X-ray positions are $\lesssim 10.3$ and $\lesssim 12.3$
kpc respectively, which agree with the observed offset
distribution. Assuming a fiducial redshift of $z=0.5$, GRB\,110112A would be
located $29 \pm 3$ kpc away from the closest and most probable host
galaxy, but this association is much less definitive ($P_{cc}(<\delta
R) \approx 0.43$) than previous host-less bursts \citep{ber10}.
Imaging with the {\it Hubble Space Telescope} may enable the
detection of a faint coincident host. These offsets are in contrast to
long GRBs which have relatively small offsets of $\approx 1$ kpc
\citep{bkd02,fls+06}. From afterglow observations, the inferred
densities for these three events may span a wide range, $n_0 \sim
10^{-4}-1$ cm$^{-3}$, while long GRBs have values of $n_0 \gtrsim
0.1$ cm$^{-3}$ \citep{sbk+06}.

The host galaxies of GRBs\,100625A and 101219A have stellar
populations that span the observed distribution of short GRB
hosts. With $\tau \approx 25-100$ Myr and log($M_*/M_{\odot}) \approx
9.1$, GRB\,101219A is at the low end of both the short GRB age and
mass distributions \citep{lb10}. This host also has one of the most
vigorous star formation rates reported for a short GRB host to date
\citep{ber09,pmm+12,bzl+12}, and an appreciable extinction of $A_V^{\rm
host}\gtrsim 2$ mag. These characteristics match more closely with the
median parameters observed for long GRB host galaxies
\citep{chg+04,wbp07,lb10}. However, an independent study based on the
$\gamma$-ray properties report a probability that GRB\,101219A is not
a collapsar of $94\%$ \citep{bnp+12}. Compared to other early-type
hosts, GRB\,100625A has a similar age ($0.6-0.8$ Gyr) and SFR limit
\citep{bpp+06,ber09,lb10,fbc+11}, but its stellar mass,
log($M_*/M_{\odot}) \approx 9.7$, is the lowest by an order of
magnitude \citep{lb10}.

\section{Host Galaxy Demographics} 

\begin{deluxetable*}{lcccccc}
\tabletypesize{\scriptsize}
\tablecolumns{7}
\tablewidth{0pc}
\tablecaption{Short GRB Host Galaxy Morphologies
\label{tab:dem}}
\tablehead {
\colhead {GRB}                    &
\colhead {$T_{90}^{a}$ }     &
\colhead {$z^{b}$}                    &
\colhead {Type$^{c}$}             &
\colhead {$90\%$ XRT uncert.$^{d}$}    &
\colhead {$P_{cc}(<\delta R)$} &
\colhead {References} \\
\colhead {}                    &
\colhead {(s)}     &
\colhead {}                    &
\colhead {}    &
\colhead {(arcsec)}             &
\colhead {}             &
\colhead {}
}
\startdata
\multicolumn{7}{c}{{\it Sub-arcsecond localized}}  \\
\noalign{\smallskip}
\hline
\noalign{\smallskip}
050709  & $0.07$ / $130$ & $0.161$  & L   & & $3 \times 10^{-3}$ & $1-3$    \\
050724A  & $3$            & $0.257$  & E   & & $2 \times 10^{-5}$ & $4-5$    \\
051221A & $1.4$          & $0.546$  & L   & & $5 \times 10^{-5}$ & $6-7$    \\
060121  & $2.0$          & $<4.1$   & ?   & & $2 \times 10^{-3}$       & $8-9$               \\
060313  & $0.7$          & $<1.7$     & ?   & & $3 \times 10^{-3}$         & $10-11$              \\
061006  & $0.4$ / $130$  & $0.4377$ & L   & & $4 \times 10^{-4}$ & $12-15$ \\
061201  & $0.8$          & $0.111$  & H/L & & \nod/$0.08$ & $9$, $16-17$ \\
070429B & $0.5$          & $0.9023$ & L   & & $3 \times 10^{-3}$ & $18-19$ \\
070707  & $1.1$          & $<3.6$   & ?   & & $7 \times 10^{-3}$                    & $20-21$ \\
070714B & $2.0$ / $64$   & $0.9224$ & L   & & $5 \times 10^{-3}$ & $19$, $22-23$ \\
070724A & $0.4$          & $0.457$  & L   & & $8 \times 10^{-4}$ & $24-25$ \\
070809  & $1.3$          & $0.473$  & H/E & & \nod/$0.03$ & $9$, $26$ \\
071227  & $1.8^{e}$      & $0.381$  & L   & & $0.01$ & $27-29$ \\
080503  & $0.3$ / $170$  & $<4.2$     & H/? & & \nod/$0.1$ & $9$, $30-31$ \\
080905A & $1.0$          & $0.1218$ & L   & & $0.01$ & $32-33$ \\
081226A & $0.4$          & $<4.1$     & ?   & &  $0.01$                   & $34-35$ \\
090305  & $0.4$          & $<4.1$     & H/? & & \nod/$0.06$ & $9$, $36$ \\
090426A & $1.3$          & $2.609$  & L   & & $1.5 \times 10^{-4}$ & $37-38$ \\
090510  & $0.3$          & $0.903$  & L   & & $8 \times 10^{-3}$ & $39-40$ \\
090515  & $0.04$         & $0.403$  & H/E & & \nod/$0.15$ & $9$, $41$ \\
091109B & $0.3$          & $<4.4$     & ? & & \nod & $42-43$ \\
100117A  & $0.3$          & $0.915$  & E   & & $7 \times 10^{-5}$ & $44-45$ \\
110112A & $0.5$          & $<5.3$     & H/?  & & $0.43$ & $46$, This work \\
111020A$^{f}$  & $0.4$    & \nod     & ?   & & $0.01$ & $47-48$ \\
111117A$^{fg}$ & $0.5$    & $1.3$    & L   & & $0.02$ & $49-50$ \\
\noalign{\smallskip}
\hline
\noalign{\smallskip}
\multicolumn{7}{c}{{\it XRT only}} \\
\noalign{\smallskip}
\hline
\noalign{\smallskip}
050509B & $0.04$          & $0.225$    & E   & $3.8$ & $5 \times 10^{-3}$ & $51-52$              \\
050813$^{h}$  & $0.6$     & $0.72/1.8$ & E/? & $2.9$ & \nod               & $53-57$ \\
051210  & $1.3$           & $>1.4$     & ?   & $1.6$ & $0.04$             & $14$, $58$               \\
060502B & $0.09$          & $0.287$    & E   & $5.2$ & $0.03$             & $59-60$             \\
060801  & $0.5$           & $1.130$    & L   & $1.5$ & $0.02$             & $61-62$ \\
061210  & $0.2$ / $85$    & $0.4095$   & L   & $3.9$ & $0.02$             & $14$, $63$ \\
061217  & $0.2$           & $0.827$    & L   & $5.5$ & $0.24^{i}$         & $14$, $64$  \\
070729$^{g}$  & $0.9$     & $0.8$      & E   & $2.5$ & $0.05$             & $65-66$ \\
080123  & $0.4$ / $115$   & $0.495$    & L   & $1.7$ & $0.004$            & $67-68$ \\
100206A  & $0.1$           & $0.4075$   & L   & $3.3$ & $0.02$             & $69-70$ \\
100625A  & $0.3$           & $0.452$    & E   & $1.8$ & $0.04$             & $71$, This work \\
101219A & $0.6$           & $0.718$    & L   & $1.7$ & $0.06$             & $72$, This work
\enddata
\tablecomments{
$^{a}$ \swift\ $15-150$ keV. For bursts with extended emission, both the duration of the
prompt spike and the duration including extended emission are
reported. \\
$^{b}$ Upper limits on redshift are based on the detection of the UV/optical afterglow and therefore the lack of suppression blueward of the Lyman limit ($\lambda_0=912$\,\AA) or Lyman-$\alpha$ line ($\lambda_0=1216$\,\AA). \\
$^{c}$ L=late-type, E=early-type, ?=inconclusive type, H=``host-less''. For each host-less burst, we also list the type of the galaxy with the lowest $P_{cc}$ (\citealt{ber10} and this work). \\
$^{d}$ Only listed for XRT bursts. \citep{gtb+07,ebp+09} \\ 
$^{e}$ Evidence at the $4\sigma$ level for extended emission is reported to $\delta t \approx 100$ s. \\
$^{f}$ Bursts with no optical afterglow, localized by \chandra. \\
$^{g}$ Bursts with galaxy type classifications based on extensive broad-band photometry \citep{lb10,mbf+12}. In particular, the host of GRB\,070729 has an inferred age ($\approx 0.98$ Gyr) and stellar mass ($\approx 4 \times 10^{10}\,M_{\odot}$; \citealt{lb10}) more consistent with an early-type designation. \\
$^{h}$ There exists disagreement in the literature regarding the association of GRB 050813 with an early-type cluster galaxy at $z=0.72$ \citep{gcn3801,gcn3808,pbc+06} or a high redshift cluster at $z=1.8$ \citep{ber06}; thus, we only display this burst for completeness but do not include it in our demographics. \\
$^{i}$ Despite the relatively high $P_{cc}$, all surrounding galaxies have $P_{cc}$ of order unity \citep{bfp+07}. \\
{\bf References}: (1) \citealt{vlr+05}; (2) \citealt{ffp+05}; (3)
\citealt{hwf+05}; (4) \citealt{gcn3667}; (5) \citealt{bpc+05}; (6)
\citealt{gcn4365}; (7) \citealt{sbk+06}; (8) \citealt{dcg+06}; (9)
\citealt{ber10}; (10) \citealt{gcn4873}; (11) \citealt{rvp+06}; (12) \citealt{gcn5717} (13)
\citealt{gcnr6}; (14) \citealt{bfp+07}; (15) \citealt{dmc+09}; (16)
\citealt{gcnr18}; (17) \citealt{sdp+07}; (18) \citealt{gcnr51}; (19)
\citealt{cbn+08}; (20) \citealt{gcn6607}; (21) \citealt{pdc+08}; (22)
\citealt{gcn6637}; (23) \citealt{gcnr70}; (24) \citealt{gcnr74}; (25)
\citealt{bcf+09}; (26) \citealt{gcnr80}; (27) \citealt{gcn7148}; (28)
\citealt{dfp+07}; (29) \citealt{snu+07}; (30) \citealt{gcnr138}; (31)
\citealt{pmg+09}; (32) \citealt{gcnr162}; (31) \citealt{rwl+10}; (34)
\citealt{gcn8735}; (35) \citealt{nkg+12}; (36) \citealt{gcn8936}; (37)
\citealt{adp+09}; (38) \citealt{lbb+10}; (39) \citealt{gcnr218}; (40)
\citealt{mkr+10}; (41) \citealt{gcn9364}; (40) \citealt{gcnr259}; (43)
\citealt{gcn10154}; (44) \citealt{gcnr269}; (45) \citealt{fbc+11};
(46) \citealt{gcn11557}; (47) \citealt{gcn12464}; (48)
\citealt{fbm+12}; (49) \citealt{sta+12}; (50) \citealt{mbf+12}; (51)
\citealt{gso+05}; (52) \citealt{bpp+06}; (53) \citealt{gcn3793}; (54)
\citealt{gcn3801}; (55) \citealt{gcn3808}; (56) \citealt{ber06}; (57)
\citealt{pbc+06}; (58) \citealt{lmf+06}; (59) \citealt{gcn5064}; (60)
\citealt{bpc+07}; (61) \citealt{gcn5381}; (62) \citealt{ber09}; (63)
\citealt{gcnr20}; (64) \citealt{gcnr21}; (65) \citealt{gcn6681}; (66)
\citealt{lb10}; (67) \citealt{gcn7223}; (68) \citealt{gcnr111}; (69)
\citealt{gcnr271}; (70) \citealt{pmm+11}; (71) \citealt{gcnr289}; (72)
\citealt{gcn11467}
}
\end{deluxetable*}

\begin{figure}
\centering
\includegraphics[angle=0,width=1.6in]{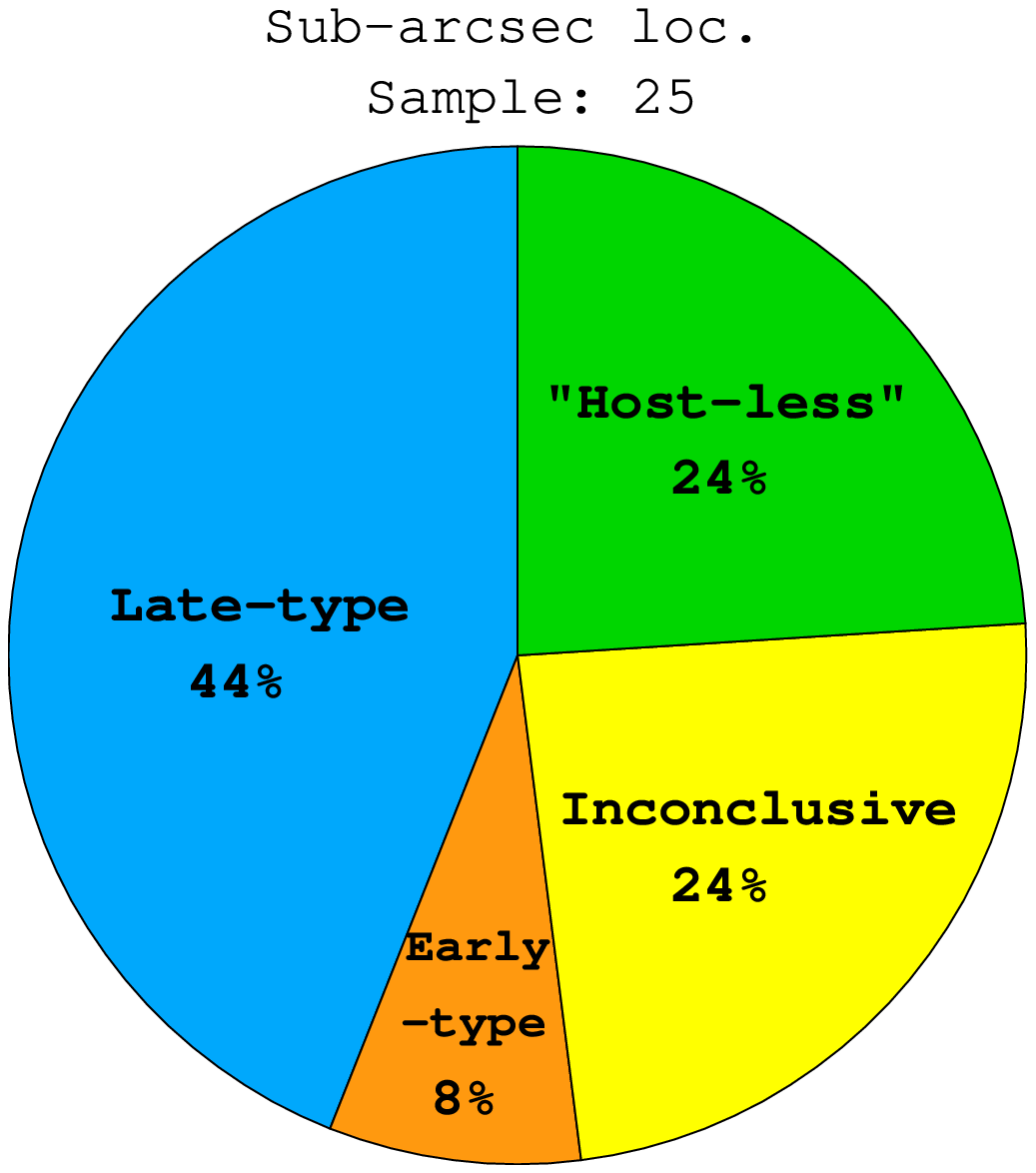}
\includegraphics[angle=0,width=1.6in]{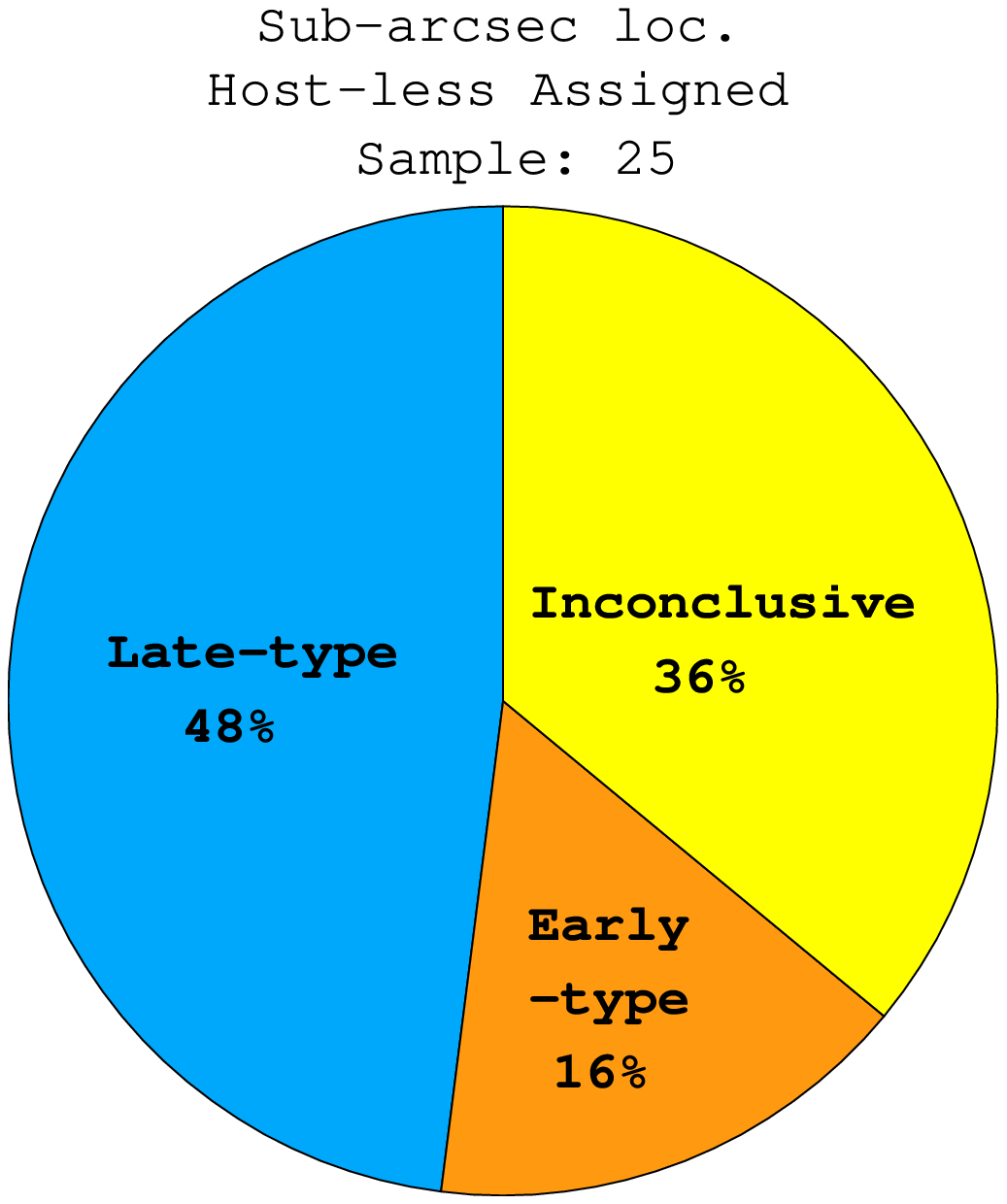} \\
\includegraphics[angle=0,width=1.6in]{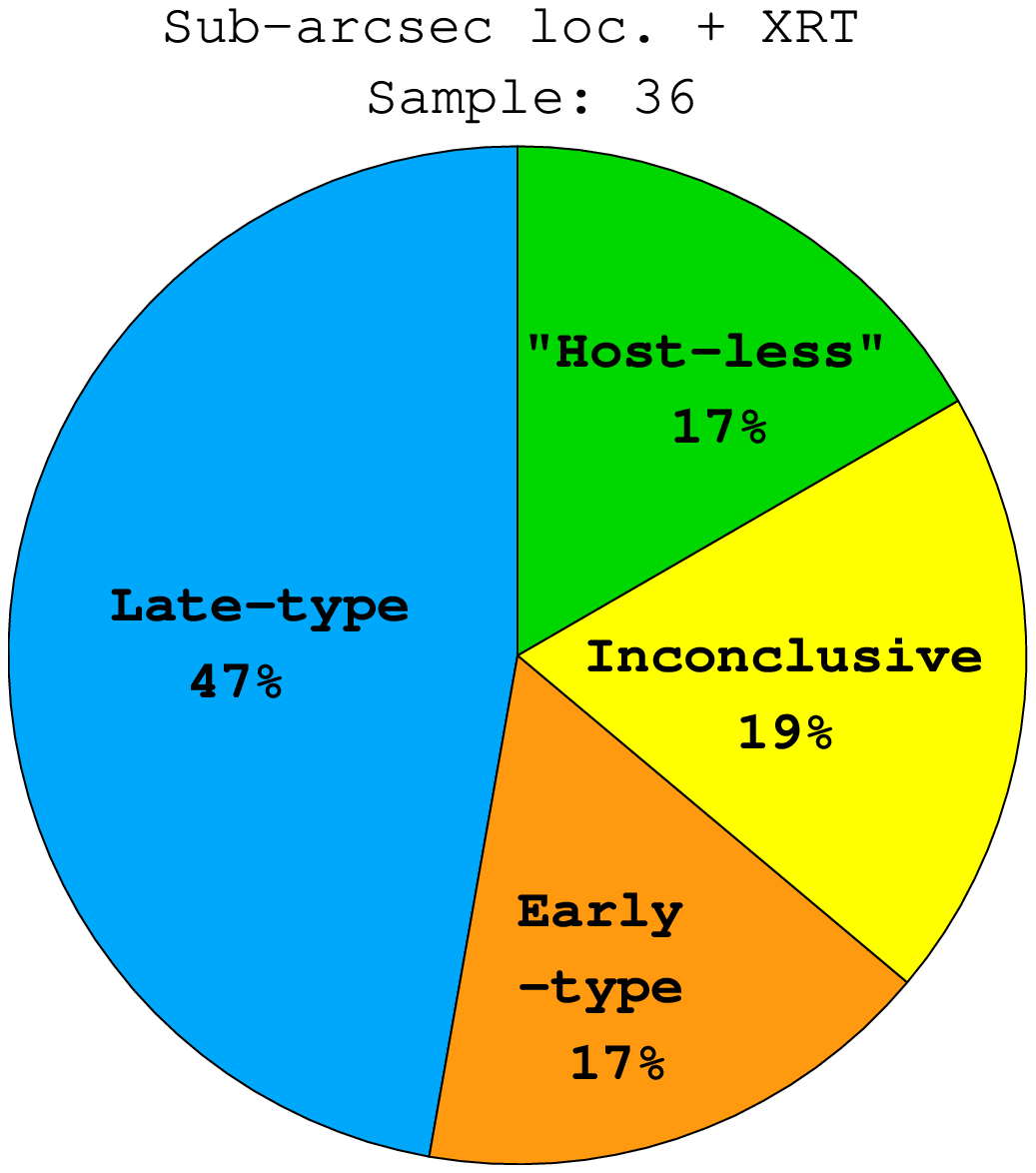}
\includegraphics[angle=0,width=1.6in]{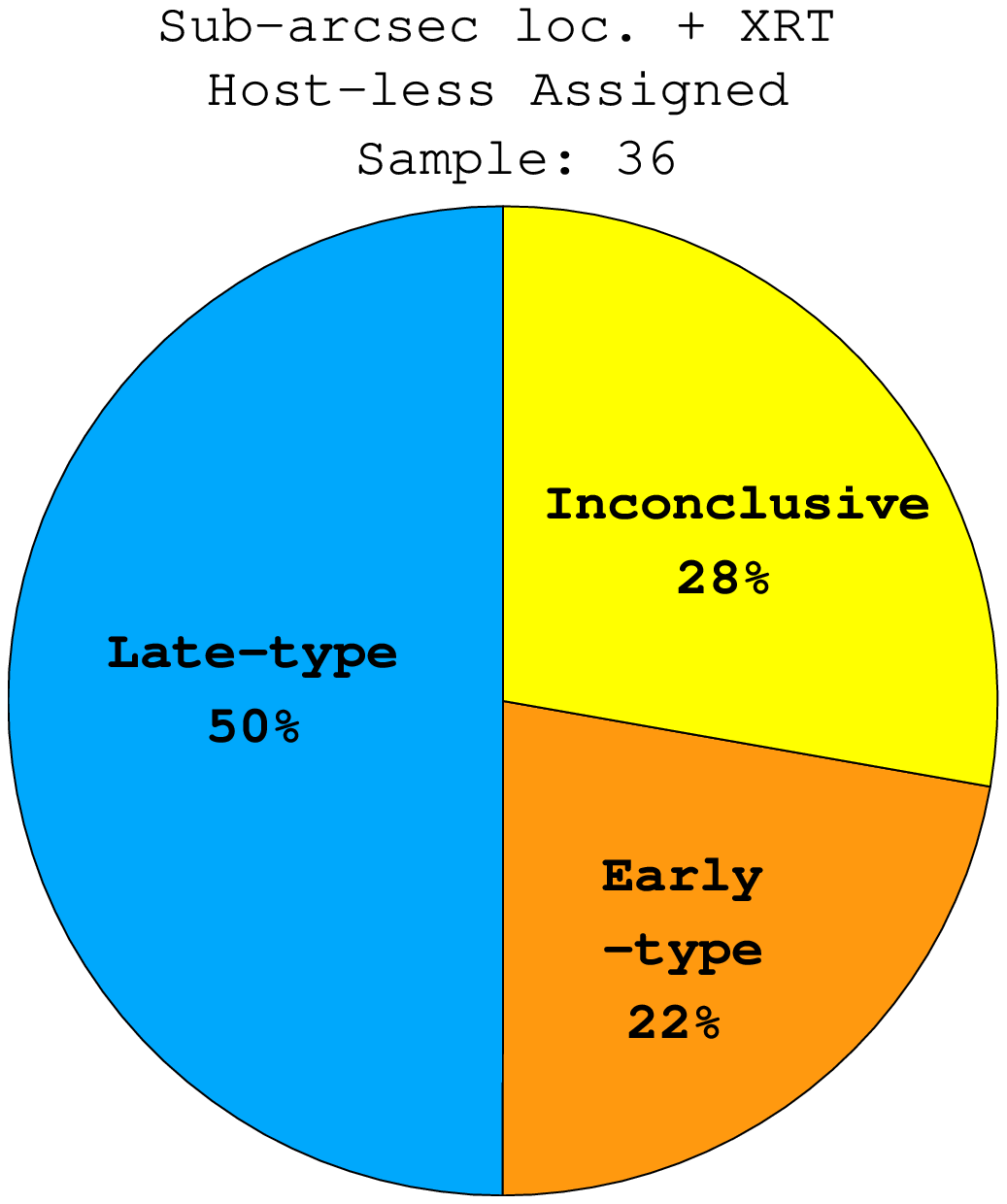} \\
\includegraphics[angle=0,width=1.6in]{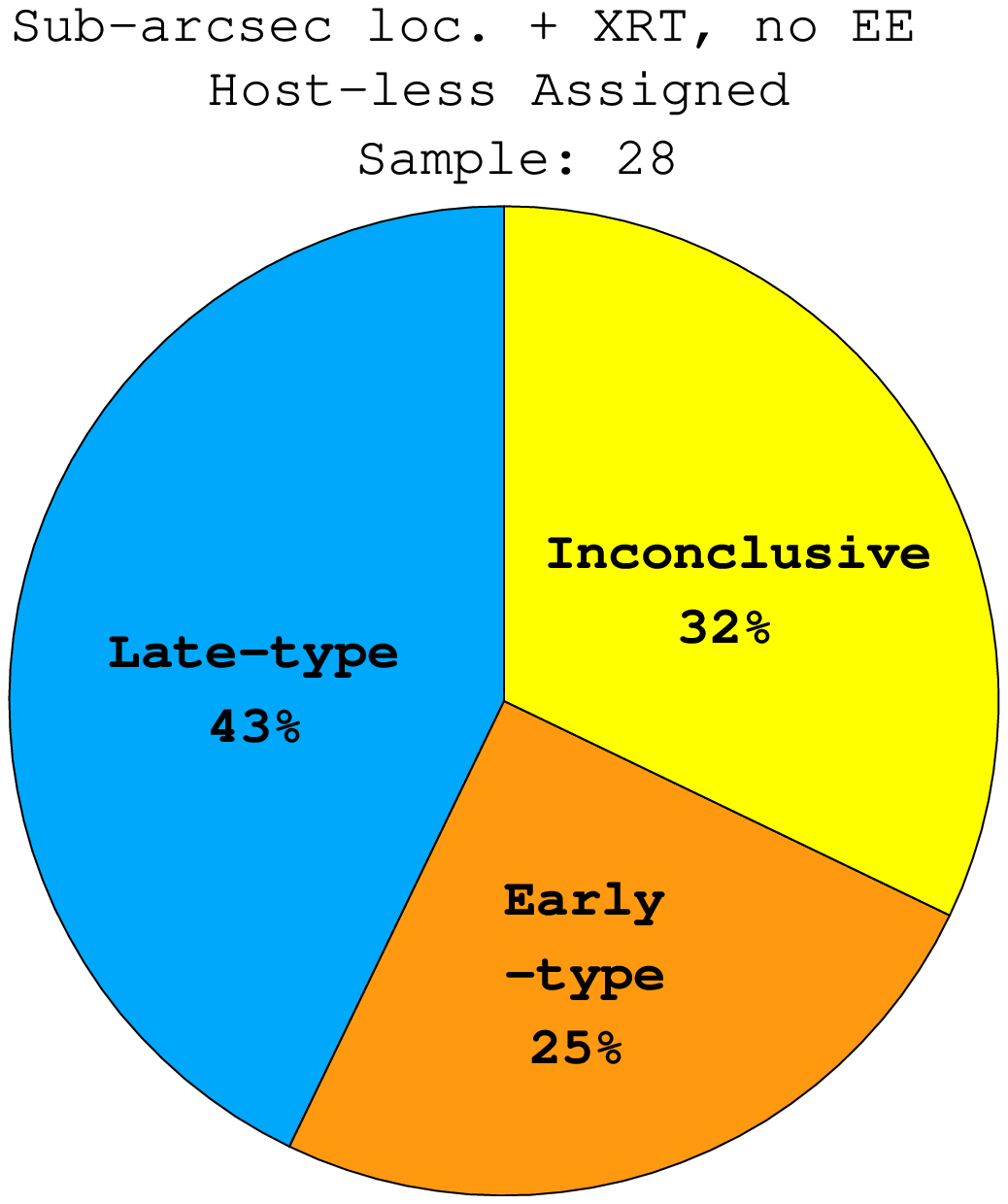}
\includegraphics[angle=0,width=1.6in]{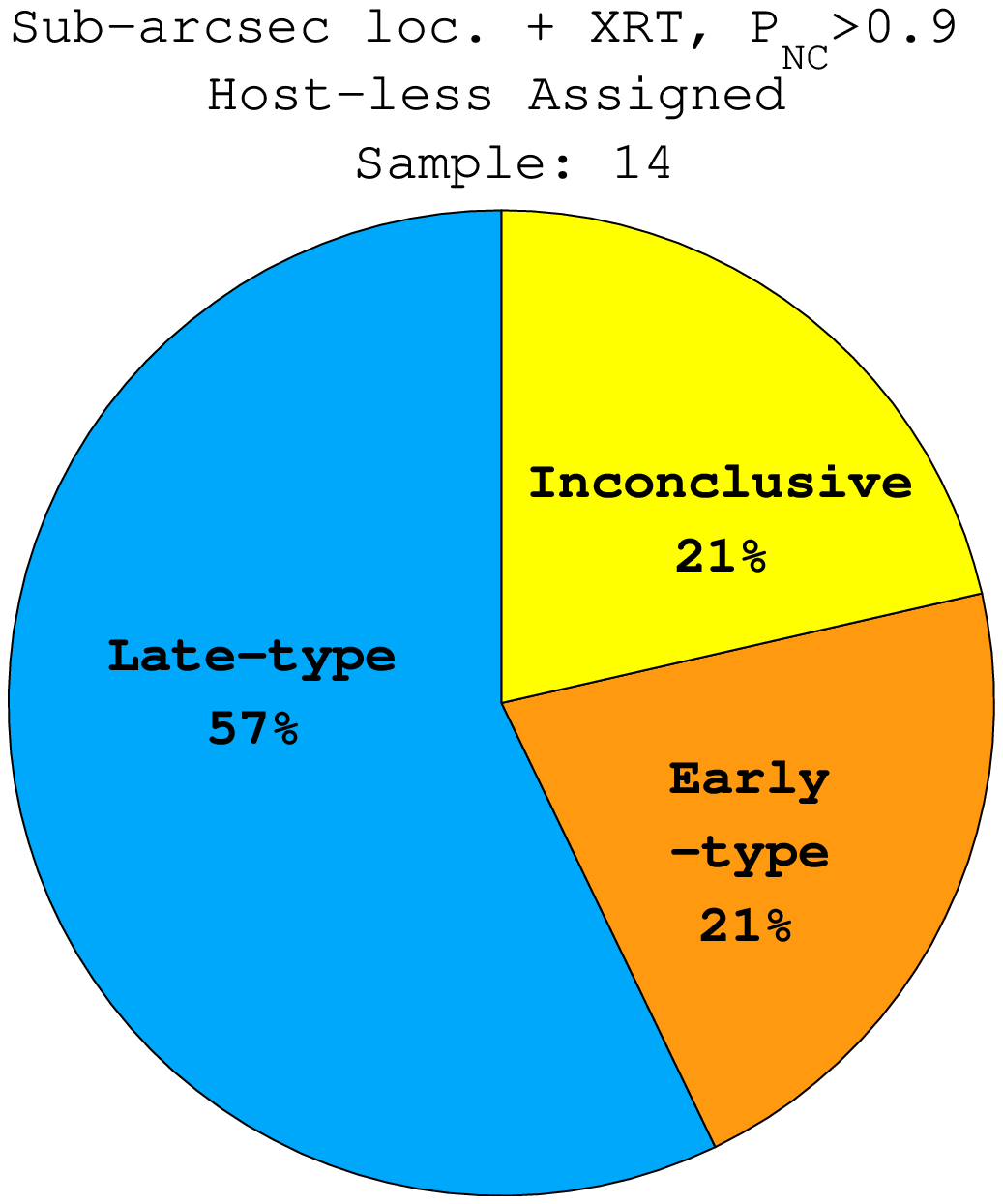}
\caption{Distribution of short GRB environments,
according to Table~\ref{tab:dem}. The fractions of late-type (blue),
early-type (orange), host-less (green) and inconclusive (yellow)
environments are shown. {\it Top:} The distribution of 25 short GRBs
with sub-arcscond localization are divided into all four categories
(left), and the 6 host-less bursts are each assigned to their most
probable host galaxy (right; \citealt{ber10} and this work). {\it
Middle:} Our full sample, including 11 short GRBs with XRT
localizations and probable hosts, is divided into all four categories
(left), and with the 6 host-less bursts assigned (right). {\it
Bottom:} Distribution of our sample for which there is no
evidence for extended emission (left) and for which $P_{\rm NC}>0.9$
(right; \citealt{bnp+12}).
\label{fig:pie}}
\end{figure}

To broadly determine and utilize the short GRB host population, we
expand upon the observations presented here and investigate the
demographics of the bulk of the \swift\ short GRB sample, quantifying the
fractions of events that explode in different types of
environments. We divide the population into four host galaxy
categories: late-type, early-type, inconclusive (coincident hosts that
are too faint to classify as late- or early-type), and ``host-less''
(lack of coincident hosts to $\gtrsim 26$ mag). All late- and
early-type designations are based on spectroscopic classification,
with the exception of two hosts, GRBs\,070729 and 111117A, which are
based on well-sampled broad-band photometry (Table~\ref{tab:dem};
\citealt{lb10,mbf+12}).

We then use our classifications to examine the relative rates of short
GRBs detected in early- and late-type galaxies. In the absence of
observational selection effects, if the overall short GRB rate tracks
stellar mass alone, the relative detection rates in early- and
late-type galaxies should match the distribution of stellar mass,
which is roughly equal at $z\sim0$ \citep{kpf+01,bmk+03,dal+07} and
shows little evolution to $z \sim 1$ \citep{isl+10}. On the other
hand, if the short GRB rate depends on a combination of stellar mass
and star formation, as in the case of Type Ia supernovae
\citep{slp+06}, we expect a distribution skewed toward star-forming
galaxies, with a late-to-early-type ratio of $>$1:1.

\subsection{Environment Fractions}

We first analyze the subset of bursts with sub-arcsecond localization
because they have the most unambiguous associations. Of the $68$ short
GRBs detected with \swift\footnotemark[12]\footnotetext[12]{We note that two
of the bursts in our sample, GRBs\,050709 and 060121, were first
discovered by the High Energy Transient Explorer~2 (HETE-2)
satellite.} as of May 2012, there are $25$ such events
(Table~\ref{tab:dem}), $2$ of which have been localized with \chandra\
(GRB\,111020A: \citealt{fbm+12}; GRB\,111117A;
\citealt{mbf+12,sta+12}), an alternative route to sub-arcsecond
positions in the absence of an optical afterglow. This population is
divided as follows: 11 ($44\%$) originate in late-type galaxies, 2
($8\%$) are in early-type galaxies, 6 ($24\%$) have hosts of
inconclusive type, and 6 ($24\%$) are host-less (\citealt{ber10} and
this work; Figure~\ref{fig:pie} and Table~\ref{tab:dem2}). From
probability of chance coincidence arguments, we can assign the
$6$ host-less GRBs to a most probable host galaxy. \citet{ber10}
investigated $5$ events, finding $2$ which likely originated in
early-type hosts (GRBs\,070809 and 090515), $1$ with a late-type host
(GRB\,061201), and $2$ with hosts of inconclusive type (GRBs\,080503
and 090305). We have shown that the remaining host-less burst, GRB
110112A lacks an obvious host galaxy (\S\ref{sec:110112a_cc}),
and we classify it as inconclusive.

Accounting for these host-less
assignments in the distribution of galaxy types, we do not find a
substantial change in the relative fractions
(Figure~\ref{fig:pie}). Considering the 16 bursts with definitive host
types, the late-to-early-type ratio is 3:1 which deviates from the
expected 1:1 distribution if the short GRB rate depends only on
stellar mass. Using binomical statistics, we test the null hypothesis
of a distribution that is intrinsically 1:1 and find that the observed
ratio has a $p$-value of only $0.04$, indicating that the null
hypothesis is disfavored (Table~\ref{tab:dem2}).

\begin{deluxetable*}{l|cccccccc}
\tabletypesize{\scriptsize}
\tablecolumns{9}
\tablewidth{0pc}
\tablecaption{Short GRB Environment Distributions
\label{tab:dem2}}
\tablehead {
\colhead {Sample}         &
\colhead {Late-type}             &
\colhead {Early-type}          &
\colhead {Inconclusive}        &
\colhead {Host-less}       &
\colhead {Total}    &
\colhead {L:E ratio$^{a}$} & 
\colhead {$P_{\rm binom}(\geq$L:E)$^{b}$} &
\colhead {Reject 1:1 distribution?$^{c}$}
}
\startdata
Sub-arcsec.                            & 11 ($44\%$) & 2 ($8\%$)  & 6 ($24\%$) & 6 ($24\%$) & 25 & 5.5:1 & $0.01$ & Yes \\
Sub-arcsec., Host-less assigned        & 12 ($48\%$) & 4 ($16\%$) & 9 ($36\%$) &            & 25 & 3:1   & $0.04$ & Yes, marginal \\
Sub-arcsec. + XRT                      & 17 ($47\%$) & 6 ($17\%$) & 7 ($19\%$) & 6 ($17\%$) & 36 & 2.8:1 & $0.02$ & Yes \\
Sub-arcsec. + XRT, Host-less assigned  & 18 ($50\%$) & 8 ($22\%$) & 10 ($28\%$) &           & 36 & 2.3:1 & $0.04$ & Yes, marginal \\
Sub-arcsec. + XRT, All Inc. are Early-type & 18 ($50\%$) & 18 ($50\%$)  &    &     & 36 & 1:1 & $0.5$ & No  \\
Sub-arcsec. + XRT, EE excluded         & 12 ($43\%$) & 7 ($25\%$) & 9 ($32\%$) &            & 28  & 1.7:1 & $0.19$ & No \\
Sub-arcsec. + XRT, $P_{\rm NC}>0.9$    & 8 ($58\%$) & 3 ($21\%$) & 3 ($21\%$) &     &   14 & 2.7:1 & $0.11$ & No
\enddata
\tablecomments{
$^{a}$ Late-to-early-type ratio \\
$^{b}$ $p$ value for finding greater than or equal to the observed L:E ratio from a 1:1 binomial distribution. \\
$^{c}$ Assumes a significance level of $0.05$.}
\end{deluxetable*}

Because the optical afterglow brightness depends on the circumburst
density, $n_0$ \citep{gs02}, the requirement of an optical afterglow
for precise positions (with the exception of the two bursts localized by
\chandra) may affect the relative rates of short
GRB detection in early- and late-type hosts if there is a correlation
between average density and galaxy type. To assess this
potential effect, we broaden our analysis to include bursts with a
single probable host galaxy ($P_{cc}(<\delta R) \lesssim 0.05$)
within or on the outskirts of XRT error circles. This sample comprises $11$
additional events\footnotemark[13]\footnotetext[13]{We exclude GRB\,050813
from our sample; see Table~\ref{tab:dem2}.} with localizations of
$1.5-5.5''$ in radius ($90\%$ containment; Table~\ref{tab:dem}),
bringing the total sample size to $36$ bursts. Since we require
sub-arcsecond localization for a burst to be classified as host-less,
the relative fraction of these events is artificially diluted by the
addition of bursts with XRT positions
(Figure~\ref{fig:pie}).

Assigning the host-less bursts to their most
probable host galaxies, we recover a similar distribution to the
sub-arcsecond localized sample: $\approx 50\%$ late-type, $\approx
20\%$ early-type, and $\approx 30\%$ inconclusive,
(Figure~\ref{fig:pie} and Table~\ref{tab:dem2}). Based on the $26$
bursts with early- and late-type designations, this gives a
late-to-early-type ratio of 2.3:1 and a low $p$-value of 0.04 for the
null hypothesis that this distribution is drawn from an intrinsically
1:1 distribution. To directly compare this 2.3:1 ratio to the 3:1
observed ratio for sub-arcsecond localized bursts, we compute the
probability of obtaining a ratio $\leq$2.3:1 from a population with a
true ratio of 3:1 using Monte Carlo simulations for the binomial
distribution. In $10^5$ trials, we calculate a high probability
of $0.82$, suggesting that there is no bias to the environment
fractions when analyzing only sub-arcsecond localized bursts.

Next, we address the remaining population of $32$ \swift\ short GRBs
excluded from the discussion thus far. The majority, $80\%$,
are affected by observing constraints that are dependent on factors completely
decoupled from any intrinsic properties of the bursts: $15$ had
\swift\ re-pointing constraints (Sun or Moon) and thus have only
$\gamma$-ray positions, $7$ have XRT positions that are highly
contaminated (in the direction of the Galactic plane or near a
saturated star, e.g. GRB\,100702A, see Appendix), and $4$ have XRT
afterglows but so far lack adequate optical/NIR follow-up to determine
the presence of a host galaxy; thus, we cannot currently distinguish
between a faint coincident host and a host-less origin for these $4$
bursts. The remaining $20\%$ (6 events) have no XRT localization
despite rapid \swift\ re-pointing ($\delta t \lesssim 2$ min), but
have a low median fluence of $f_\gamma \approx 2 \times 10^{-8}$ erg
cm$^{-2}$ compared to the rest of the population with $\langle
f_{\gamma,{\rm SGRB}}\rangle \approx 2 \times 10^{-7}$ erg cm$^{-2}$
($15-150$ keV; Figure~\ref{fig:ft}). Therefore, the lack of detectable
emission with XRT may be related to an intrinsically lower energy
scale. In summary, we do not expect the exclusion of these $32$ bursts
to have a substantial effect on the relative morphological fractions.

The low observed early-type fraction is likely attributed to one of two
possibilities: (1) it is more challenging to identify early-type
galaxies at higher redshifts, and thus a disproportionate fraction of
the bursts designated as inconclusive are in fact early-type; or (2)
short GRBs preferentially occur in late-type galaxies due to the
intrinsic properties of their progenitors.

We explore the former option by investigating the inconclusive
population in more detail. Spectral energy distributions of early-type
galaxies generally lack strong emission lines, and the most prominent
features, the $4000$ \AA\ break and the \ion{Ca}{2} H\&K absorption
lines, are redshifted out of the range of most optical spectrographs
for $z\gtrsim 1.5$, making spectroscopic identifications particularly
difficult at these redshifts. However, more effective studies
selecting for distant early-type field galaxies by their photometric
optical/NIR colors detect a nearly constant number of early-types
between $z\approx 1-1.5$ \citep{sdp+04}, with a typical AB color of
$1-4$ mag, depending on the choice of optical/NIR filters
\citep{sdp+04,to04}. Of the $10$ inconclusive host galaxies, $4$ have
optical/NIR color information but yield only poor constraints of
$\lesssim 3-5$ mag due to NIR non-detections and faint optical
magnitudes, and $5$ lack reported NIR observations. The only
inconclusive host galaxy with multi-band detections, GRB\,060121, has
$R-H \approx 2.4$ mag; however, the optical afterglow and objects in
the vicinity are comparably red, suggesting a $z>2$ origin as an
explanation for the red host color \citep{ltf+06}. $K$-band imaging to
depths of $\gtrsim 23$ AB mag might enable progress in deducing what
fraction of the inconclusive population is more likely early-type. To
set an extreme upper bound on the true early-type fraction, if we
assume that all inconclusive hosts are early-types, the projected
early-type fraction is $\sim 50\%$ (Table~\ref{tab:dem2}).

We now turn to the second option, that short GRBs preferentially
originate from late-type galaxies. While the predicted demographics of
NS-NS/NS-BH merger populations are currently not well-constrained
\citep{bpb+06}, we can use the observed short GRB population to assess the
implications for the progenitors. We expect to find roughly equal
early- and late-type fractions if stellar mass is the sole parameter
determining the short GRB rate. However, we only observe this for $z<0.4$
(6 events; Figure~\ref{fig:zhist}). For $z>0.4$, the late-type fraction is
consistently higher, with a late-to-early-type ratio of $\gtrsim$2:1. These
results, along with the previous finding that the short GRB rate per
unit stellar mass is $2-5$ times higher in late-type hosts
\citep{lb10}, suggest that the short GRB rate is dependent upon a
combination of stellar mass and star formation. In the context of
NS-NS/NS-BH mergers, if the delay times of the systems which give
rise to short GRBs are very long ($\gtrsim$ few Gyr), we would expect
a dominant population of early-type hosts at $z \sim 0$. Instead, the
current demographics show a preference for late-type
galaxies. Along with the inferred stellar population ages from SED
modeling \citep{lb10}, this suggests moderate delay times of
$\lesssim$ few Gyr. For a delay time distribution of the form $P(\tau)
\propto \tau^{n}$, this translates to $n \lesssim -1$. We note that
this result is similar to Type Ia supernovae which have $n \approx
-1.1$ \citep{msg10,mmb12}, and is in contrast to previous short GRB
results which claimed substantially longer average delay times of
$\sim 4-8$ Gyr for lognormal lifetime distributions based on smaller
numbers of events \citep{ngf06,zr07,gno+08}.

In summary, we find that unless all inconclusive hosts are early-type,
the short GRB host distribution is skewed toward late-type galaxies, with the
most likely ranges for the early- and late-type fractions of $\approx
20-40\%$ and $\approx 60-80\%$, respectively, for the entire short GRB
population. Furthermore, for most cuts on the sample we find that the
null hypothesis of a 1:1 distribution can be mildly or strongly
rejected.

\subsection{Comparison with $\gamma$-ray Properties}

\begin{figure}
\centering
\includegraphics*[angle=0,width=3.2in]{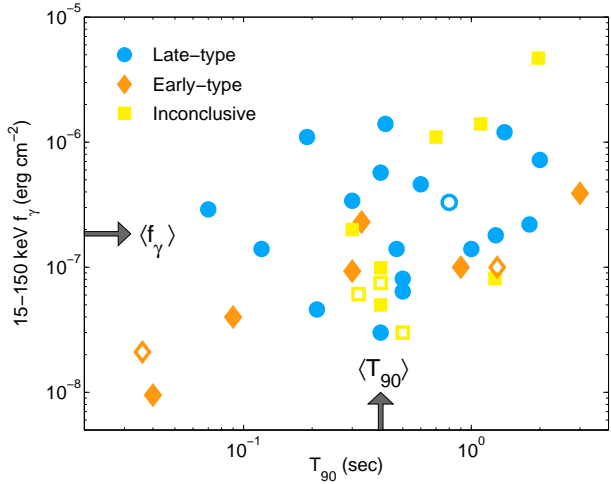}
\caption{Fluence, $f_{\gamma}$, ($15-150$ keV) versus duration,
$T_{90}$ for the sub-arcsecond localized + XRT sample of $36$ \swift\
short GRBs. Bursts are classified by morphological type
(Table~\ref{tab:dem}) as late-type (blue), early-type (orange) and
inconclusive (yellow). Open symbols denote host-less
assignments. The median $f_\gamma \approx 2 \times 10^{-7}$ erg
cm$^{-2}$ and $T_{90} \approx 0.4$ s are labeled. The majority of
events have $f_{\gamma}\approx 10^{-8}-10^{-6}$ erg cm$^{-2}$.
\label{fig:ft}}
\end{figure}

We next investigate whether there is contamination in our sample
from collapsars by analyzing trends between morphological type and
$\gamma$-ray properties. We find that bursts in early- and late-type
galaxies span the entire distribution of observed $T_{90}$ for short
GRBs, with a median value of $0.4$ s (Figure~\ref{fig:ft}). Using a
Kolmogorov-Smirnov (K-S) test, we find that the two populations are
consistent with being drawn from the same underlying distribution
($p=0.43$). The claim becomes stronger when we compare the combined
early-type and inconclusive distribution with the late-type
distribution ($p=0.94$). On the other hand, the corresponding K-S
tests for the fluence distributions (Figure~\ref{fig:ft}) yield
marginal $p$-values of $0.05$, suggesting that bursts associated with
early- and late-types may not be drawn from the same underlying
distribution in $f_{\gamma}$.

A recent study by \citet{bnp+12} used the $\gamma$-ray properties
($T_{90}$ and spectral hardness) to derive a probability that each event
is {\it not} a collapsar ($P_{\rm NC}$), excluding $8$ bursts which
have reported evidence for extended emission. Of the $29$ bursts that overlap in our
samples, $14$ have a high probability of not arising from a
collapsar ($P_{\rm NC}>0.9$). If these probabilities are robust, and
there is contamination from collapsars in our full sample, we would expect the galaxy
type fractions for the population with $P_{\rm NC}>0.9$ to differ from
the overall sample. In particular, by including only
high-probability non-collapsar events, we would presumably be
excluding mostly late-type galaxies since all long GRBs/collapsars are
found in star-forming galaxies. Therefore, one would naively expect
the late-to-early-type ratio to {\it decrease} with respect to the
full sample. However, we find that the late-to-early-type ratio for
this sample is 2.7:1 (Table~\ref{tab:dem2}; Figure~\ref{fig:pie})
which is higher than the 2.3:1 ratio inferred for the sample of $36$
short GRBs.

However, $P_{\rm NC}$ values are not reported for bursts with extended
emission. Thus, for a more direct comparison, we evaluate the subset
of $28$ short GRBs without extended emission (Figure~\ref{fig:pie}),
and calculate a late-to-early-type ratio of 1.7:1
(Table~\ref{tab:dem2}). Interestingly, all bursts with extended emission
originate in late-type (or inconclusive) galaxies, with the exception
of GRB\,050724A. Since the ratio for the $P_{\rm NC}>0.9$ population
is more skewed toward late-type galaxies with 2.7:1, the probability of
obtaining a $\geq$2.7:1 ratio in 14 events from an intrinsically 1:7:1
distribution is moderate, $0.37$. This not only demonstrates no noticeable
contamination to the short GRB host type distribution when including
bursts with reportedly high probabilities of being collapsars, but
also calls into question the reliability or importance of these probabilities in
assessing the true population of short GRBs.

\section{Conclusions}

We present broad-band observations of three short GRBs: GRB 100625A
associated with an early-type galaxy at
$z=0.452$, GRB 101219A associated with an active star-forming galaxy
at $z=0.718$, and GRB\,110112A which has a sub-arcsecond localization
from an optical afterglow but no coincident host galaxy to deep
optical limits, and no convincing putative host within $5^{\circ}$ of the
burst location. These observations showcase the diversity of short GRB
environments and give direct clues to the nature of the short GRB
progenitor: the moderate physical offsets and low inferred densities
can be interpreted as evidence for a compact binary progenitor.

We also undertake the first comprehensive study of host demographics for the
full \swift\ short GRB population, classifying bursts by their
host galaxy type. We emphasize several key conclusions:

\begin{enumerate}
\item The sample of sub-arcsecond localized bursts have a host galaxy
distribution of $\approx 50\%$ late-type, $\approx 20\%$ early-type
and $\approx 30\%$ of inconclusive type after assigning host-less
bursts. The inclusion of bursts with \swift/XRT positions and
convincing host associations ($P_{cc}(<\delta R) \lesssim 0.05$) does
not affect the relative fractions.

\item The observed late-to-early-type ratio is $\gtrsim$2:1, and most
cuts to the sample demonstrate that an intrinsically 1:1 distribution
is improbable. The only way to obtain equal fractions with the
observed events is by assuming that all inconclusive hosts are
early-type galaxies at $z \gtrsim 1$.

\item The most likely ranges for the early- and late-type fractions
are $\approx 20-40\%$ and $\approx 60-80\%$. The preference toward
late-type galaxies suggests that both stellar mass and star formation
play roles in determining the short GRB rate. Furthermore, in the context of the
NS-NS/NS-BH mergers, the observed short GRB population is not
dominated by systems with very long delay times, but instead with typical delay times of $\lesssim$ few Gyr.

\item There is no clear trend between $T_{90}$ and host
galaxy type, while there may be a relationship between $f_{\gamma}$
and host type. When excluding the population of bursts reported to be
likely collapsars ($>90\%$ probability), the late-type fraction
increases relative to the overall short GRB sample, suggesting that
these probabilities are not reliable in assessing the true population.

\end{enumerate}

Looking forward, our study has demonstrated that detailed observations of short
GRB afterglows and environments hold the key to understanding the
underlying population of progenitors. In particular, we emphasize the
importance of deep NIR observations to determine the early-type
fraction within the inconclusive population of hosts, and {\it Hubble
Space Telescope} observations of short GRBs which lack coincident host
galaxies to ground-based optical limits ($\approx 26$ mag). A
concerted analysis of broad-band short GRB afterglows would complement
this study by providing constraints on the basic properties of the
bursts (i.e., energy scale, circumburst density), and help to
determine whether there are any correlations between these basic
properties and galactic environment. Finally,
constraints on theoretical predictions for the relative fractions of
early- and late-type galaxies which host NS-NS/NS-BH mergers and their
delay time distributions will enable a direct comparison to the
observed short GRB population.

\appendix

\section{GRB\,100628A}

GRB\,100628A was detected by {\it Swift}/BAT and the Anti-Coincidence
System on INTEGRAL on 2010 June 28.345 UT with $T_{90}=0.036 \pm 0.009$
s ($15-350$ keV), $f_\gamma = 2.5 \pm 0.5 \times
10^{-8}$ erg cm$^{-2}$ ($15-150$ keV), and peak energy $E_{\rm peak} =
74.1 \pm 11.4$ keV. The ground-calculated position is
RA=\ra{15}{03}{46.2}, Dec=\dec{-31}{39}{10.2} with an uncertainty of
$2.1'$ \citep{gcnr290}.

\subsection{X-ray Observations}

XRT began observing the field at $\delta t=86$ s and detected an X-ray
source in coincidence with the core of a bright galaxy. The lack of
fading of this source confirmed by {\it Chandra}/ACIS-S observations at $\delta
t=4.4$ days suggests an AGN origin
\citep{gcnr290,gcn10942}. Furthermore, we use binomial statistics and
a $10$-pixel region centered on the source to calculate the
probability of a chance fluctuation, finding a high probability of
$15\%$. Thus, this source is ruled out as the afterglow of
GRB\,100628A. A second candidate afterglow was reported based on $7$
counts over $3.8$ ks in the time interval $\delta t=92-7200$ s, which
translates to a count rate of $0.0017^{+0.0008}_{-0.0006}$ counts
s$^{-1}$ ($0.3-10$ keV; \citealt{gcnr290}). UVOT, which commenced
observations at $\delta t=90$ s, did not detect a coincident source to
$\gtrsim 20.2$ mag ($white$ filter; \citealt{gcnr290}).

We re-analyze the same time interval of XRT data and use the {\tt
ximage} routine in the HEASOFT package to measure the significance of
the source. In a blind search, we find the source has a significance
of $2.3\sigma$. Late-time XRT and {\it Chandra} observations confirm
that the source has faded by a factor of $\sim 15$ from the claimed
initial X-ray flux \citep{gcn10942}. However, we do not include this
burst in our sample of short GRBs with XRT positions due to the low
significance of the initial source. We caution against classifying this
burst as XRT-localized in future short GRB samples.

\appendix 
\label{100702}
\section{GRB\,100702A}

{\it Swift}/BAT detected GRB\,100702A on 2010 July 02.044 UT  with
$T_{90}=0.16 \pm 0.03$ s ($15-350$ keV) and $f_{\gamma} = (1.2 \pm 0.1)
\times 10^{-7}$ erg cm$^{-2}$ ($15-150$ keV) at a ground-calculated
position of RA=\ra{16}{22}{46.4} and Dec=$-$\dec{56}{32}{57.4} with an
uncertainty of $1.4'$ in radius \citep{gcnr292}.

\subsection{X-ray Observations}

XRT started observing the field at $\delta t=94$ s and identified a
fading X-ray counterpart with a final UVOT-enhanced positional
accuracy of $2.4''$ (Table~\ref{tab:info};
\citealt{gtb+07,ebp+09}). UVOT commenced observations at $\delta
t=101$ s and no source was identified in the $white$ filter to a limit
of $\gtrsim 18$ mag \citep{gcnr292}. The XRT light curve is best fit
with a broken power law with decay indices of $\alpha_{X,1} =
-0.86^{+0.17}_{-0.24}$ and $\alpha_{X,2} = -5.04^{+0.34}_{-0.37}$, and
a break time at $\delta t=202$ s \citep{ebp+09}.

We extract a spectrum from the XRT data (method described in
\S\ref{sec:100625prompt}) and utilize the full PC data set,
where there is no evidence for spectral evolution. Our best-fit model
is characterized by $\Gamma=2.7 \pm 0.3$ and $(4.4 \pm 2.0) \times
10^{21}$ cm$^{-2}$ in excess of the substantial Galactic value,
$N_{\rm H,MW}=2.8 \times 10^{21}$ cm$^{-2}$ \citep{kbh+05}. We note
that the burst is in the direction of the Galactic Center
($b=-4.8^{\circ}$) and therefore the uncertainties on $N_{\rm H,MW}$
are likely larger than the typical $10\%$. Our results are consistent
with the automatic fits by \citet{ebp+09}.

\subsection{Optical/NIR Observations and Afterglow Limits}

\begin{figure}
\centering
\includegraphics[angle=0,width=6.4in]{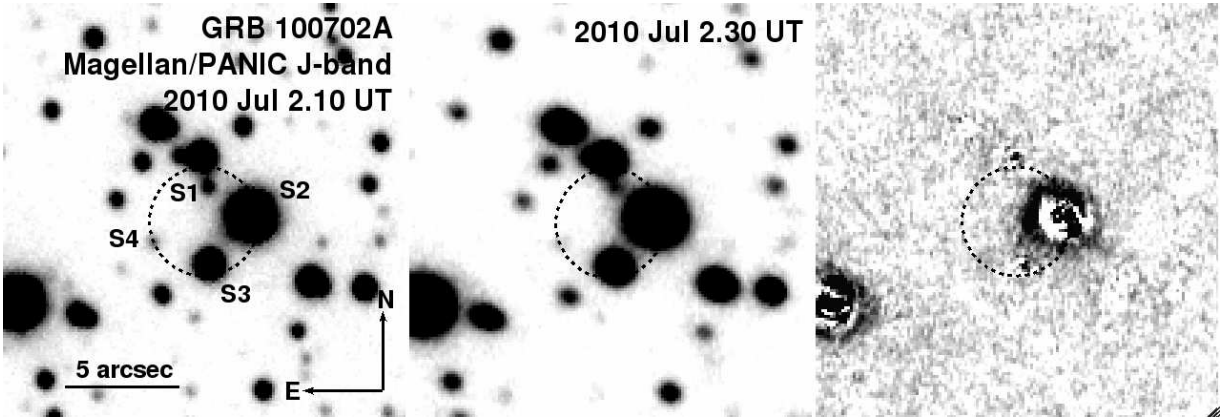}
\caption{Magellan/PANIC $J$-band observations of the host galaxy of
GRB\,100702A. The XRT error circle has a radius of $2.4''$ ($90\%$
containment; black). {\it
Left:} $\delta t=1.3$ hr. {\it Center:} $\delta t=6.1$ hr. {\it
Right:} Digital image subtraction of the two epochs reveals no
afterglow to a $3\sigma$ limit of $J \gtrsim 23.3$ mag.
\label{fig:100702_J}}
\end{figure}

We obtained $J$-band observations of the field of GRB\,100702A with
PANIC at $\delta t=1.3$ hr (Figure~\ref{fig:100702_J}). We detect $4$
sources within or near the outskirts of the XRT error circle (S1-S4 in
Figure~\ref{fig:100702_J}). S2 and S3 have stellar PSFs, while S1 and S4
have non-stellar PSFs. Previously reported $J$-band observations also
confirm that S2 and S3 are stars \citep{nkg+12}, while S1 and S4 have
not been reported in the literature\footnotemark[14]\footnotetext[14]{Our
PANIC observations show that source ``C'' in \citet{nkg+12} is
actually three blended sources, including S1. The remaining two
sources are outside of the XRT error circle.}  To assess any
fading within the XRT position, we obtained a second set of $J$-band
observations at $\delta t=6.1$ hr. Digital image subtraction reveal no
residuals to a $3\sigma$ limit of $J \gtrsim 23.3$ mag
(Table~\ref{tab:obs}). We caution that this limit only applies to
$2/3$ of the error circle due to contamination from the saturated
star, S2 (Figure~\ref{fig:100702_J}). In addition, we obtained
$i$-band observations with IMACS at $\delta t=247.3$ days and we
do not detect any additional sources in or around the XRT error
circle (Table~\ref{tab:obs}).

\subsection{Probabilities of Chance Coincidence}

We calculate $P_{cc}(<\delta R)$ for S1 and S4 to assess either source
as a putative host galaxy for GRB\,100702A. Source S1 is fully inside
the XRT error circle while S4 lies on the outskirts of the XRT error
circle. We perform PSF photometry for both sources
(Table~\ref{tab:obs}), and calculate their probabilities of chance
coincidence: $P_{cc}(<\delta R) \approx 0.02$ for S1 and $P(<\delta R)
\approx 0.04$ for S2 using the $3\sigma$ XRT position radius of
$4.5''$. This analysis suggests that either source is a likely host
for GRB\,100702A, and we cannot currently distinguish which is more
likely. We also note that the significant contamination makes it
difficult to exclude the possibility that there is a brighter galaxy
within the XRT error circle. Therefore, we do not include GRB\,100702A
in our sample of bursts with XRT localization, and consider this field
to have observing constraints which prevent more in-depth analysis.

\acknowledgments

\noindent We thank F.~di Mille for observing on behalf of the Berger
GRB group at Harvard. The Berger GRB group is supported by the
National Science Foundation under Grant AST-1107973, and by NASA/Swift
AO6 grant NNX10AI24G and A07 grant NNX12AD69G. Partial support was
also provided by the National Aeronautics and Space Administration
through Chandra Award Number GO1-12072X issued by the Chandra X-ray
Observatory Center, which is operated by the Smithsonian Astrophysical
Observatory for and on behalf of the National Aeronautics Space
Administration under contract NAS8-03060. This paper includes data
gathered with the 6.5 meter Magellan Telescopes located at Las
Campanas Observatory, Chile.  This work is based in part on
observations obtained at the Gemini Observatory, which is operated by
the Association of Universities for Research in Astronomy, Inc., under
a cooperative agreement with the NSF on behalf of the Gemini
partnership: the National Science Foundation (United States), the
Science and Technology Facilities Council (United Kingdom), the
National Research Council (Canada), CONICYT (Chile), the Australian
Research Council (Australia), Ministério da Ciência, Tecnologia e
Inovação (Brazil) and Ministerio de Ciencia, Tecnología e Innovación
Productiva (Argentina). This work made use of data supplied by the UK
Swift Science Data Centre at the University of Leicester. SBC
acknowledges generous support from Gary \& Cynthia Bengier, the
Richard \& Rhoda Goldman Fund, the Christopher R. Redlich Fund, the
TABASGO Foundation, and NSF grants AST-0908886 and AST-1211916


\begin{thebibliography}{}

\bibitem[\protect\citeauthoryear{{Alard}}{{Alard}}{2000}]{ala00}
{Alard}, C. 2000, \aaps, 144, 363

\bibitem[\protect\citeauthoryear{{Anderson} et~al.}{{Anderson}
  et~al.}{2012}]{ahj+12}
{Anderson}, J.~P., {Habergham}, S.~M., {James}, P.~A.,  \& {Hamuy}, M. 2012,
  \mnras, 424, 1372

\bibitem[\protect\citeauthoryear{{Antonelli} et~al.}{{Antonelli}
  et~al.}{2009}]{adp+09}
{Antonelli}, L.~A., et~al. 2009, \aap, 507, L45

\bibitem[\protect\citeauthoryear{{Barthelmy} et~al.}{{Barthelmy}
  et~al.}{2009}]{gcn9364}
{Barthelmy}, S.~D., et~al. 2009, GRB Coordinates Network, 9364, 1

\bibitem[\protect\citeauthoryear{{Barthelmy}, {Sakamoto}, \&
  {Stamatikos}}{{Barthelmy} et~al.}{2011}]{gcn11557}
{Barthelmy}, S.~D., {Sakamoto}, T.,  \& {Stamatikos}, M. 2011, GRB Coordinates
  Network, 11557, 1

\bibitem[\protect\citeauthoryear{{Belczynski} et~al.}{{Belczynski}
  et~al.}{2006}]{bpb+06}
{Belczynski}, K., {Perna}, R., {Bulik}, T., {Kalogera}, V., {Ivanova}, N.,  \&
  {Lamb}, D.~Q. 2006, \apj, 648, 1110

\bibitem[\protect\citeauthoryear{{Bell} et~al.}{{Bell} et~al.}{2003}]{bmk+03}
{Bell}, E.~F., {McIntosh}, D.~H., {Katz}, N.,  \& {Weinberg}, M.~D. 2003,
  \apjs, 149, 289

\bibitem[\protect\citeauthoryear{{Berger}}{{Berger}}{2005}]{gcn3801}
{Berger}, E. 2005, GRB Coordinates Network, 3801, 1

\bibitem[\protect\citeauthoryear{{Berger}}{{Berger}}{2006}]{ber06}
{Berger}, E. 2006, in American Institute of Physics Conference Series, Vol.
  836, Gamma-Ray Bursts in the Swift Era, ed. S.~S. {Holt}, N.~{Gehrels}, \&
  J.~A. {Nousek}, 33

\bibitem[\protect\citeauthoryear{{Berger}}{{Berger}}{2009}]{ber09}
{Berger}, E. 2009, \apj, 690, 231

\bibitem[\protect\citeauthoryear{{Berger}}{{Berger}}{2010a}]{ber10}
{Berger}, E. 2010a, \apj, 722, 1946

\bibitem[\protect\citeauthoryear{{Berger}}{{Berger}}{2010b}]{gcn10942}
{Berger}, E. 2010b, GRB Coordinates Network, 10942, 1

\bibitem[\protect\citeauthoryear{{Berger} et~al.}{{Berger}
  et~al.}{2009}]{bcf+09}
{Berger}, E., {Cenko}, S.~B., {Fox}, D.~B.,  \& {Cucchiara}, A. 2009, \apj,
  704, 877

\bibitem[\protect\citeauthoryear{{Berger} et~al.}{{Berger}
  et~al.}{2007}]{bfp+07}
{Berger}, E., et~al. 2007, \apj, 664, 1000

\bibitem[\protect\citeauthoryear{{Berger} et~al.}{{Berger}
  et~al.}{2005}]{bpc+05}
{Berger}, E., et~al. 2005, \nat, 438, 988

\bibitem[\protect\citeauthoryear{{Berger} et~al.}{{Berger}
  et~al.}{2012}]{bzl+12}
{Berger}, E., et~al. 2012, ArXiv e-prints

\bibitem[\protect\citeauthoryear{{Bhat}}{{Bhat}}{2010}]{gcn10912}
{Bhat}, P.~N. 2010, GRB Coordinates Network, 10912, 1

\bibitem[\protect\citeauthoryear{{Bloom}, {Kulkarni}, \& {Djorgovski}}{{Bloom}
  et~al.}{2002}]{bkd02}
{Bloom}, J.~S., {Kulkarni}, S.~R.,  \& {Djorgovski}, S.~G. 2002, \aj, 123, 1111

\bibitem[\protect\citeauthoryear{{Bloom} et~al.}{{Bloom} et~al.}{2007}]{bpc+07}
{Bloom}, J.~S., et~al. 2007, \apj, 654, 878

\bibitem[\protect\citeauthoryear{{Bloom} et~al.}{{Bloom} et~al.}{2006}]{bpp+06}
{Bloom}, J.~S., et~al. 2006, \apj, 638, 354

\bibitem[\protect\citeauthoryear{{Breeveld} \& {Stamatikos}}{{Breeveld} \&
  {Stamatikos}}{2011}]{gcn11560}
{Breeveld}, A.~A.,  \& {Stamatikos}, M. 2011, GRB Coordinates Network, 11560, 1

\bibitem[\protect\citeauthoryear{{Bromberg} et~al.}{{Bromberg}
  et~al.}{2012}]{bnp+12}
{Bromberg}, O., {Nakar}, E., {Piran}, T.,  \& {Sari}, R. 2012, ArXiv e-prints

\bibitem[\protect\citeauthoryear{{Bruzual} \& {Charlot}}{{Bruzual} \&
  {Charlot}}{2003}]{bc03}
{Bruzual}, G.,  \& {Charlot}, S. 2003, \mnras, 344, 1000

\bibitem[\protect\citeauthoryear{{Burrows} et~al.}{{Burrows}
  et~al.}{2006}]{bgc+06}
{Burrows}, D.~N., et~al. 2006, \apj, 653, 468

\bibitem[\protect\citeauthoryear{{Burrows} et~al.}{{Burrows}
  et~al.}{2005}]{bhn+05}
{Burrows}, D.~N., et~al. 2005, \ssr, 120, 165

\bibitem[\protect\citeauthoryear{{Cannizzo} et~al.}{{Cannizzo}
  et~al.}{2006}]{gcnr20}
{Cannizzo}, J.~K., et~al. 2006, GCNR, 20, 1 (2006), 20, 1

\bibitem[\protect\citeauthoryear{{Cardelli}, {Clayton}, \& {Mathis}}{{Cardelli}
  et~al.}{1989}]{ccm89}
{Cardelli}, J.~A., {Clayton}, G.~C.,  \& {Mathis}, J.~S. 1989, \apj, 345, 245

\bibitem[\protect\citeauthoryear{{Castro-Tirado} et~al.}{{Castro-Tirado}
  et~al.}{2005}]{cdg+05}
{Castro-Tirado}, A.~J., et~al. 2005, \aap, 439, L15

\bibitem[\protect\citeauthoryear{{Cenko} et~al.}{{Cenko} et~al.}{2008}]{cbn+08}
{Cenko}, S.~B., et~al. 2008, ArXiv e-prints, 802

\bibitem[\protect\citeauthoryear{{Chapman} et~al.}{{Chapman}
  et~al.}{2008}]{clw+08}
{Chapman}, R., {Levan}, A.~J., {Wynn}, G.~A., {Davies}, M.~B., {King}, A.~R.,
  {Priddey}, R.~S.,  \& {Tanvir}, N.~R. 2008, in American Institute of Physics
  Conference Series, Vol. 983, 40 Years of Pulsars: Millisecond Pulsars,
  Magnetars and More, ed. {C.~Bassa, Z.~Wang, A.~Cumming, \& V.~M.~Kaspi}, 301

\bibitem[\protect\citeauthoryear{{Christensen}, {Hjorth}, \&
  {Gorosabel}}{{Christensen} et~al.}{2004}]{chg+04}
{Christensen}, L., {Hjorth}, J.,  \& {Gorosabel}, J. 2004, \aap, 425, 913

\bibitem[\protect\citeauthoryear{{Cummings} et~al.}{{Cummings}
  et~al.}{2005}]{gcn4365}
{Cummings}, J., et~al. 2005, GRB Coordinates Network, 4365, 1

\bibitem[\protect\citeauthoryear{{D'Avanzo} et~al.}{{D'Avanzo}
  et~al.}{2007}]{dfp+07}
{D'Avanzo}, P., {Fiore}, F., {Piranomonte}, S., {Covino}, S., {Tagliaferri},
  G., {Chincarini}, G.,  \& {Stella}, L. 2007, GRB Coordinates Network, 7152, 1

\bibitem[\protect\citeauthoryear{{D'Avanzo} et~al.}{{D'Avanzo}
  et~al.}{2009}]{dmc+09}
{D'Avanzo}, P., et~al. 2009, \aap, 498, 711

\bibitem[\protect\citeauthoryear{{de Pasquale}, {Markwardt}, \&
  {Sbarufatti}}{{de Pasquale} et~al.}{2010}]{gcnr269}
{de Pasquale}, M., {Markwardt}, C.,  \& {Sbarufatti}, B. 2010, GCN Report, 269,
  1

\bibitem[\protect\citeauthoryear{{de Ugarte Postigo} et~al.}{{de Ugarte
  Postigo} et~al.}{2006}]{dcg+06}
{de Ugarte Postigo}, A., et~al. 2006, \apjl, 648, L83

\bibitem[\protect\citeauthoryear{{Djorgovski} et~al.}{{Djorgovski}
  et~al.}{1998}]{dkb+98}
{Djorgovski}, S.~G., {Kulkarni}, S.~R., {Bloom}, J.~S., {Goodrich}, R.,
  {Frail}, D.~A., {Piro}, L.,  \& {Palazzi}, E. 1998, \apjl, 508, L17

\bibitem[\protect\citeauthoryear{{Driver} et~al.}{{Driver}
  et~al.}{2007}]{dal+07}
{Driver}, S.~P., {Allen}, P.~D., {Liske}, J.,  \& {Graham}, A.~W. 2007, \apjl,
  657, L85

\bibitem[\protect\citeauthoryear{{Eichler} et~al.}{{Eichler}
  et~al.}{1989}]{elp+89}
{Eichler}, D., {Livio}, M., {Piran}, T.,  \& {Schramm}, D.~N. 1989, \nat, 340,
  126

\bibitem[\protect\citeauthoryear{{Evans} et~al.}{{Evans} et~al.}{2009}]{ebp+09}
{Evans}, P.~A., et~al. 2009, \mnras, 397, 1177

\bibitem[\protect\citeauthoryear{{Evans} et~al.}{{Evans}
  et~al.}{2011}]{gcn11556}
{Evans}, P.~A., {Goad}, M.~R., {Osborne}, J.~P.,  \& {Beardmore}, A.~P. 2011,
  GRB Coordinates Network, 11556, 1

\bibitem[\protect\citeauthoryear{{Foley}, {Bloom}, \& {Chen}}{{Foley}
  et~al.}{2005}]{gcn3808}
{Foley}, R.~J., {Bloom}, J.~S.,  \& {Chen}, H.-W. 2005, GRB Coordinates
  Network, 3808, 1

\bibitem[\protect\citeauthoryear{{Fong} et~al.}{{Fong} et~al.}{2011}]{fbc+11}
{Fong}, W., et~al. 2011, \apj, 730, 26

\bibitem[\protect\citeauthoryear{{Fong}, {Berger}, \& {Fox}}{{Fong}
  et~al.}{2010}]{fbf10}
{Fong}, W., {Berger}, E.,  \& {Fox}, D.~B. 2010, \apj, 708, 9

\bibitem[\protect\citeauthoryear{{Fong} et~al.}{{Fong} et~al.}{2012}]{fbm+12}
{Fong}, W.-f., et~al. 2012, ArXiv e-prints

\bibitem[\protect\citeauthoryear{{Fox} et~al.}{{Fox} et~al.}{2005}]{ffp+05}
{Fox}, D.~B., et~al. 2005, \nat, 437, 845

\bibitem[\protect\citeauthoryear{{Fruchter} et~al.}{{Fruchter}
  et~al.}{2006}]{fls+06}
{Fruchter}, A.~S., et~al. 2006, \nat, 441, 463

\bibitem[\protect\citeauthoryear{{Fryer}, {Woosley}, \& {Hartmann}}{{Fryer}
  et~al.}{1999}]{fwh99}
{Fryer}, C.~L., {Woosley}, S.~E.,  \& {Hartmann}, D.~H. 1999, \apj, 526, 152

\bibitem[\protect\citeauthoryear{{Gal-Yam} et~al.}{{Gal-Yam}
  et~al.}{2008}]{gno+08}
{Gal-Yam}, A., et~al. 2008, \apj, 686, 408

\bibitem[\protect\citeauthoryear{{Gehrels} et~al.}{{Gehrels}
  et~al.}{2004}]{ggg+04}
{Gehrels}, N., et~al. 2004, \apj, 611, 1005

\bibitem[\protect\citeauthoryear{{Gehrels} et~al.}{{Gehrels}
  et~al.}{2005}]{gso+05}
{Gehrels}, N., et~al. 2005, \nat, 437, 851

\bibitem[\protect\citeauthoryear{{Gelbord} et~al.}{{Gelbord}
  et~al.}{2010}]{gcn11461}
{Gelbord}, J.~M., et~al. 2010, GRB Coordinates Network, 11461, 1

\bibitem[\protect\citeauthoryear{{Gelbord} \& {Grupe}}{{Gelbord} \&
  {Grupe}}{2010}]{gcn11474}
{Gelbord}, J.~M.,  \& {Grupe}, D. 2010, GRB Coordinates Network, 11474, 1

\bibitem[\protect\citeauthoryear{{Goad} et~al.}{{Goad} et~al.}{2007}]{gtb+07}
{Goad}, M.~R., et~al. 2007, \aap, 476, 1401

\bibitem[\protect\citeauthoryear{{Golenetskii} et~al.}{{Golenetskii}
  et~al.}{2010a}]{gcn10890}
{Golenetskii}, S., et~al. 2010a, GRB Coordinates Network, 10890, 1

\bibitem[\protect\citeauthoryear{{Golenetskii} et~al.}{{Golenetskii}
  et~al.}{2010b}]{gcn11470}
{Golenetskii}, S., et~al. 2010b, GRB Coordinates Network, 11470, 1

\bibitem[\protect\citeauthoryear{{Gotz} et~al.}{{Gotz} et~al.}{2007}]{gcn6607}
{Gotz}, D., {Beckmann}, V., {Mereghetti}, S.,  \& {Paizis}, A. 2007, GRB
  Coordinates Network, 6607, 1

\bibitem[\protect\citeauthoryear{{Granot} \& {Sari}}{{Granot} \&
  {Sari}}{2002}]{gs02}
{Granot}, J.,  \& {Sari}, R. 2002, \apj, 568, 820

\bibitem[\protect\citeauthoryear{{Grupe} et~al.}{{Grupe} et~al.}{2006}]{gbp+06}
{Grupe}, D., {Burrows}, D.~N., {Patel}, S.~K., {Kouveliotou}, C., {Zhang}, B.,
  {M{\'e}sz{\'a}ros}, P., {Wijers}, R.~A.~M.,  \& {Gehrels}, N. 2006, \apj,
  653, 462

\bibitem[\protect\citeauthoryear{{Hakobyan} et~al.}{{Hakobyan}
  et~al.}{2008}]{hpm+08}
{Hakobyan}, A.~A., {Petrosian}, A.~R., {McLean}, B., {Kunth}, D., {Allen},
  R.~J., {Turatto}, M.,  \& {Barbon}, R. 2008, \aap, 488, 523

\bibitem[\protect\citeauthoryear{{Hjorth} et~al.}{{Hjorth}
  et~al.}{2005a}]{hsg+05}
{Hjorth}, J., et~al. 2005a, \apjl, 630, L117

\bibitem[\protect\citeauthoryear{{Hjorth} et~al.}{{Hjorth}
  et~al.}{2005b}]{hwf+05}
{Hjorth}, J., et~al. 2005b, \nat, 437, 859

\bibitem[\protect\citeauthoryear{{Holland} et~al.}{{Holland}
  et~al.}{2010a}]{gcn10884}
{Holland}, S.~T., et~al. 2010a, GRB Coordinates Network, 10884, 1

\bibitem[\protect\citeauthoryear{{Holland} et~al.}{{Holland}
  et~al.}{2010b}]{gcnr289}
{Holland}, S.~T., {Landsman}, W.~B., {Page}, K.~L.,  \& {Stamatikos}, M. 2010b,
  GCN Report, 289, 1

\bibitem[\protect\citeauthoryear{{Hoversten} et~al.}{{Hoversten}
  et~al.}{2009}]{gcnr218}
{Hoversten}, E.~A., {Krimm}, H.~A., {Grupe}, D., {Kuin}, N.~P.~M., {Barthelmy},
  S.~D., {Burrows}, D.~N., {Roming}, P.,  \& {Gehrels}, N. 2009, GCN Report,
  218, 1

\bibitem[\protect\citeauthoryear{{Ilbert} et~al.}{{Ilbert}
  et~al.}{2010}]{isl+10}
{Ilbert}, O., et~al. 2010, \apj, 709, 644

\bibitem[\protect\citeauthoryear{{Immler} et~al.}{{Immler}
  et~al.}{2010}]{gcnr290}
{Immler}, S., {Starling}, R.~L.~C., {Evans}, P.~A., {Barthelmy}, S.~D.,  \&
  {Sakamoto}, T. 2010, GCN Report, 290, 1

\bibitem[\protect\citeauthoryear{{Kalberla} et~al.}{{Kalberla}
  et~al.}{2005}]{kbh+05}
{Kalberla}, P.~M.~W., {Burton}, W.~B., {Hartmann}, D., {Arnal}, E.~M.,
  {Bajaja}, E., {Morras}, R.,  \& {P{\"o}ppel}, W.~G.~L. 2005, \aap, 440, 775

\bibitem[\protect\citeauthoryear{{Kelly} \& {Kirshner}}{{Kelly} \&
  {Kirshner}}{2012}]{kk12}
{Kelly}, P.~L.,  \& {Kirshner}, R.~P. 2012, \apj, 759, 107

\bibitem[\protect\citeauthoryear{{Kennicutt}}{{Kennicutt}}{1998}]{ken98}
{Kennicutt}, R.~C., Jr. 1998, \araa, 36, 189

\bibitem[\protect\citeauthoryear{{Kochanek} et~al.}{{Kochanek}
  et~al.}{2001}]{kpf+01}
{Kochanek}, C.~S., et~al. 2001, \apj, 560, 566

\bibitem[\protect\citeauthoryear{{Kodaka} et~al.}{{Kodaka}
  et~al.}{2007}]{gcn6637}
{Kodaka}, N., et~al. 2007, GRB Coordinates Network, 6637, 1

\bibitem[\protect\citeauthoryear{{Kouveliotou} et~al.}{{Kouveliotou}
  et~al.}{1993}]{kmf+93}
{Kouveliotou}, C., {Meegan}, C.~A., {Fishman}, G.~J., {Bhat}, N.~P., {Briggs},
  M.~S., {Koshut}, T.~M., {Paciesas}, W.~S.,  \& {Pendleton}, G.~N. 1993,
  \apjl, 413, L101

\bibitem[\protect\citeauthoryear{{Krimm} et~al.}{{Krimm}
  et~al.}{2005}]{gcn3667}
{Krimm}, H., et~al. 2005, GRB Coordinates Network, 3667, 1

\bibitem[\protect\citeauthoryear{{Krimm} et~al.}{{Krimm}
  et~al.}{2008}]{gcn8735}
{Krimm}, H.~A., et~al. 2008, GRB Coordinates Network, 8735, 1

\bibitem[\protect\citeauthoryear{{Krimm} et~al.}{{Krimm}
  et~al.}{2010a}]{gcn11467}
{Krimm}, H.~A., et~al. 2010a, GRB Coordinates Network, 11467, 1

\bibitem[\protect\citeauthoryear{{Krimm} et~al.}{{Krimm}
  et~al.}{2009}]{gcn8936}
{Krimm}, H.~A., et~al. 2009, GRB Coordinates Network, 8936, 1

\bibitem[\protect\citeauthoryear{{Krimm} et~al.}{{Krimm}
  et~al.}{2010b}]{gcnr271}
{Krimm}, H.~A., {Cummings}, J.~R., {Evans}, P.~A.,  \& {Marshall}, F.~E. 2010b,
  GCN Report, 271, 1

\bibitem[\protect\citeauthoryear{{Kuin} \& {Gelbord}}{{Kuin} \&
  {Gelbord}}{2010}]{gcn11472}
{Kuin}, N.~P.~M.,  \& {Gelbord}, J.~M. 2010, GRB Coordinates Network, 11472, 1

\bibitem[\protect\citeauthoryear{{La Parola} et~al.}{{La Parola}
  et~al.}{2006}]{lmf+06}
{La Parola}, V., et~al. 2006, \aap, 454, 753

\bibitem[\protect\citeauthoryear{{Laskar}, {Berger}, \& {Chary}}{{Laskar}
  et~al.}{2011}]{lbc11}
{Laskar}, T., {Berger}, E.,  \& {Chary}, R.-R. 2011, \apj, 739, 1

\bibitem[\protect\citeauthoryear{{Le Floc'h} et~al.}{{Le Floc'h}
  et~al.}{2003}]{ldm+03}
{Le Floc'h}, E., et~al. 2003, \aap, 400, 499

\bibitem[\protect\citeauthoryear{{Leibler} \& {Berger}}{{Leibler} \&
  {Berger}}{2010}]{lb10}
{Leibler}, C.~N.,  \& {Berger}, E. 2010, \apj, 725, 1202

\bibitem[\protect\citeauthoryear{{Levan} et~al.}{{Levan}
  et~al.}{2006a}]{ltf+06}
{Levan}, A.~J., et~al. 2006a, \apjl, 648, L9

\bibitem[\protect\citeauthoryear{{Levan} et~al.}{{Levan}
  et~al.}{2009}]{gcn10154}
{Levan}, A.~J., {Tanvir}, N.~R., {Hjorth}, J., {Malesani}, D., {de Ugarte
  Postigo}, A.,  \& {D'Avanzo}, P. 2009, GRB Coordinates Network, 10154, 1

\bibitem[\protect\citeauthoryear{{Levan} et~al.}{{Levan}
  et~al.}{2006b}]{lwc+06}
{Levan}, A.~J., {Wynn}, G.~A., {Chapman}, R., {Davies}, M.~B., {King}, A.~R.,
  {Priddey}, R.~S.,  \& {Tanvir}, N.~R. 2006b, \mnras, 368, L1

\bibitem[\protect\citeauthoryear{{Levesque} et~al.}{{Levesque}
  et~al.}{2010}]{lbb+10}
{Levesque}, E.~M., et~al. 2010, \mnras, 401, 963

\bibitem[\protect\citeauthoryear{{Li} et~al.}{{Li} et~al.}{2011}]{lcl+11}
{Li}, W., {Chornock}, R., {Leaman}, J., {Filippenko}, A.~V., {Poznanski}, D.,
  {Wang}, X., {Ganeshalingam}, M.,  \& {Mannucci}, F. 2011, \mnras, 412, 1473

\bibitem[\protect\citeauthoryear{{Mannucci} et~al.}{{Mannucci}
  et~al.}{2005}]{mdp+05}
{Mannucci}, F., {Della Valle}, M., {Panagia}, N., {Cappellaro}, E., {Cresci},
  G., {Maiolino}, R., {Petrosian}, A.,  \& {Turatto}, M. 2005, \aap, 433, 807

\bibitem[\protect\citeauthoryear{{Mao} et~al.}{{Mao} et~al.}{2008}]{gcnr138}
{Mao}, J., {Guidorzi}, C., {Ukwatta}, T., {Brown}, P.~J., {Barthelmy}, S.~D.,
  {Burrows}, D.~N., {Roming}, P.,  \& {Gehrels}, N. 2008, GCN Report, 138, 1

\bibitem[\protect\citeauthoryear{{Maoz}, {Mannucci}, \& {Brandt}}{{Maoz}
  et~al.}{2012}]{mmb12}
{Maoz}, D., {Mannucci}, F.,  \& {Brandt}, T.~D. 2012, \mnras, 426, 3282

\bibitem[\protect\citeauthoryear{{Maoz}, {Sharon}, \& {Gal-Yam}}{{Maoz}
  et~al.}{2010}]{msg10}
{Maoz}, D., {Sharon}, K.,  \& {Gal-Yam}, A. 2010, \apj, 722, 1879

\bibitem[\protect\citeauthoryear{{Maraston}}{{Maraston}}{2005}]{mar05}
{Maraston}, C. 2005, \mnras, 362, 799

\bibitem[\protect\citeauthoryear{{Margutti} et~al.}{{Margutti}
  et~al.}{2012a}]{mbf+12}
{Margutti}, R., et~al. 2012a, \apj, 756, 63

\bibitem[\protect\citeauthoryear{{Margutti} et~al.}{{Margutti}
  et~al.}{2010}]{mgg+10}
{Margutti}, R., et~al. 2010, \mnras, 402, 46

\bibitem[\protect\citeauthoryear{{Margutti} et~al.}{{Margutti}
  et~al.}{2012b}]{mzb+12}
{Margutti}, R., et~al. 2012b, ArXiv e-prints

\bibitem[\protect\citeauthoryear{{Markwardt} et~al.}{{Markwardt}
  et~al.}{2006}]{gcn4873}
{Markwardt}, C., et~al. 2006, GRB Coordinates Network, 4873, 1

\bibitem[\protect\citeauthoryear{{Markwardt} et~al.}{{Markwardt}
  et~al.}{2007}]{gcnr51}
{Markwardt}, C., {Beardmore}, A., {Marshall}, F.~E., {Schady}, P., {Barthelmy},
  S.~D., {Burrows}, D.~N., {Roming}, P.,  \& {Gehrels}, N. 2007, GCN Report,
  51, 1

\bibitem[\protect\citeauthoryear{{Marshall} et~al.}{{Marshall}
  et~al.}{2006}]{gcnr18}
{Marshall}, F., {Perri}, M., {Stratta}, G., {Barthelmy}, S.~D., {Krimm}, H.,
  {Burrows}, D.~N., {Roming}, P.,  \& {Gehrels}, N. 2006, GCN Report, 18, 1

\bibitem[\protect\citeauthoryear{{Marshall} et~al.}{{Marshall}
  et~al.}{2007}]{gcnr80}
{Marshall}, F.~E., {Barthelmy}, S.~D., {Burrows}, D.~N., {Chester}, M.~M.,
  {Cummings}, J., {Evans}, P.~A., {Roming}, P.,  \& {Gehrels}, N. 2007, GCNR,
  80, 1 (2007), 80, 1

\bibitem[\protect\citeauthoryear{{McBreen} et~al.}{{McBreen}
  et~al.}{2010}]{mkr+10}
{McBreen}, S., et~al. 2010, \aap, 516, A71

\bibitem[\protect\citeauthoryear{{Metzger}, {Quataert}, \&
  {Thompson}}{{Metzger} et~al.}{2008}]{mqt08}
{Metzger}, B.~D., {Quataert}, E.,  \& {Thompson}, T.~A. 2008, \mnras, 385, 1455

\bibitem[\protect\citeauthoryear{{Naito} et~al.}{{Naito}
  et~al.}{2010}]{gcn10889}
{Naito}, H., {Sako}, T., {Suzuki}, D., {Kobara}, S., {Omori}, K., {Nagayama},
  T., {Kurita}, M.,  \& {Oi}, N. 2010, GRB Coordinates Network, 10889, 1

\bibitem[\protect\citeauthoryear{{Nakar}, {Gal-Yam}, \& {Fox}}{{Nakar}
  et~al.}{2006}]{ngf06}
{Nakar}, E., {Gal-Yam}, A.,  \& {Fox}, D.~B. 2006, \apj, 650, 281

\bibitem[\protect\citeauthoryear{{Narayan}, {Paczynski}, \& {Piran}}{{Narayan}
  et~al.}{1992}]{npp92}
{Narayan}, R., {Paczynski}, B.,  \& {Piran}, T. 1992, \apjl, 395, L83

\bibitem[\protect\citeauthoryear{{Nicuesa Guelbenzu} et~al.}{{Nicuesa
  Guelbenzu} et~al.}{2012}]{nkg+12}
{Nicuesa Guelbenzu}, A., et~al. 2012, ArXiv e-prints

\bibitem[\protect\citeauthoryear{{Oates} et~al.}{{Oates}
  et~al.}{2009}]{gcnr259}
{Oates}, S.~R., {Page}, K.~L., {Evans}, P.~A.,  \& {Markwardt}, C.~B. 2009, GCN
  Report, 259, 1

\bibitem[\protect\citeauthoryear{{Oemler} \& {Tinsley}}{{Oemler} \&
  {Tinsley}}{1979}]{ot79}
{Oemler}, A., Jr.,  \& {Tinsley}, B.~M. 1979, \aj, 84, 985

\bibitem[\protect\citeauthoryear{{Pagani} et~al.}{{Pagani}
  et~al.}{2008}]{gcnr162}
{Pagani}, C., {Racusin}, J.~L., {Brown}, P.,  \& {Cummings}, J. 2008, GCN
  Report, 162, 1

\bibitem[\protect\citeauthoryear{{Page} \& {Holland}}{{Page} \&
  {Holland}}{2010}]{gcn10888}
{Page}, K.~L.,  \& {Holland}, S.~T. 2010, GRB Coordinates Network, 10888, 1

\bibitem[\protect\citeauthoryear{{Panaitescu}, {Kumar}, \&
  {Narayan}}{{Panaitescu} et~al.}{2001}]{pkn01}
{Panaitescu}, A., {Kumar}, P.,  \& {Narayan}, R. 2001, \apjl, 561, L171

\bibitem[\protect\citeauthoryear{{Perley} et~al.}{{Perley}
  et~al.}{2009}]{pmg+09}
{Perley}, D.~A., et~al. 2009, \apj, 696, 1871

\bibitem[\protect\citeauthoryear{{Perley} et~al.}{{Perley}
  et~al.}{2011}]{pmm+11}
{Perley}, D.~A., {Modjaz}, M., {Morgan}, A.~N., {Cenko}, S.~B., {Bloom}, J.~S.,
  {Butler}, N.~R., {Filippenko}, A.~V.,  \& {Miller}, A.~A. 2011, ArXiv
  e-prints

\bibitem[\protect\citeauthoryear{{Perley} et~al.}{{Perley}
  et~al.}{2012}]{pmm+12}
{Perley}, D.~A., {Modjaz}, M., {Morgan}, A.~N., {Cenko}, S.~B., {Bloom}, J.~S.,
  {Butler}, N.~R., {Filippenko}, A.~V.,  \& {Miller}, A.~A. 2012, \apj, 758,
  122

\bibitem[\protect\citeauthoryear{{Piranomonte} et~al.}{{Piranomonte}
  et~al.}{2008}]{pdc+08}
{Piranomonte}, S., et~al. 2008, \aap, 491, 183

\bibitem[\protect\citeauthoryear{{Poole} et~al.}{{Poole} et~al.}{2008}]{pbp+08}
{Poole}, T.~S., et~al. 2008, \mnras, 383, 627

\bibitem[\protect\citeauthoryear{{Predehl} \& {Schmitt}}{{Predehl} \&
  {Schmitt}}{1995}]{ps95}
{Predehl}, P.,  \& {Schmitt}, J.~H.~M.~M. 1995, \aap, 293, 889

\bibitem[\protect\citeauthoryear{{Prochaska} et~al.}{{Prochaska}
  et~al.}{2006}]{pbc+06}
{Prochaska}, J.~X., et~al. 2006, \apj, 642, 989

\bibitem[\protect\citeauthoryear{{Qin} et~al.}{{Qin} et~al.}{1998}]{qwc+98}
{Qin}, B., {Wu}, X.-P., {Chu}, M.-C., {Fang}, L.-Z.,  \& {Hu}, J.-Y. 1998,
  \apjl, 494, L57

\bibitem[\protect\citeauthoryear{{Racusin}, {Barbier}, \& {Landsman}}{{Racusin}
  et~al.}{2007}]{gcnr70}
{Racusin}, J., {Barbier}, L.,  \& {Landsman}, W. 2007, GCNR, 70, 1 (2007), 70,
  1

\bibitem[\protect\citeauthoryear{{Roming} et~al.}{{Roming}
  et~al.}{2006}]{rvp+06}
{Roming}, P.~W.~A., et~al. 2006, \apj, 651, 985

\bibitem[\protect\citeauthoryear{{Rowlinson} et~al.}{{Rowlinson}
  et~al.}{2010}]{rwl+10}
{Rowlinson}, A., et~al. 2010, \mnras

\bibitem[\protect\citeauthoryear{{Sakamoto} et~al.}{{Sakamoto}
  et~al.}{2011}]{gcn12464}
{Sakamoto}, T., et~al. 2011, GRB Coordinates Network, 12464, 1

\bibitem[\protect\citeauthoryear{{Sakamoto} et~al.}{{Sakamoto}
  et~al.}{2007}]{snu+07}
{Sakamoto}, T., {Norris}, J., {Ukwatta}, T., {Barthelmy}, S.~D., {Gehrels}, N.,
   \& {Stamatikos}, M. 2007, GRB Coordinates Network, 7156, 1

\bibitem[\protect\citeauthoryear{{Sakamoto} et~al.}{{Sakamoto}
  et~al.}{2012}]{sta+12}
{Sakamoto}, T., et~al. 2012, ArXiv e-prints

\bibitem[\protect\citeauthoryear{{Sari}, {Piran}, \& {Halpern}}{{Sari}
  et~al.}{1999}]{sph99}
{Sari}, R., {Piran}, T.,  \& {Halpern}, J.~P. 1999, \apjl, 519, L17

\bibitem[\protect\citeauthoryear{{Sato} et~al.}{{Sato} et~al.}{2005}]{gcn3793}
{Sato}, G., et~al. 2005, GRB Coordinates Network, 3793, 1

\bibitem[\protect\citeauthoryear{{Sato} et~al.}{{Sato} et~al.}{2006a}]{gcn5064}
{Sato}, G., et~al. 2006a, GRB Coordinates Network, 5064, 1

\bibitem[\protect\citeauthoryear{{Sato} et~al.}{{Sato} et~al.}{2007a}]{gcn6681}
{Sato}, G., et~al. 2007a, GRB Coordinates Network, 6681, 1

\bibitem[\protect\citeauthoryear{{Sato} et~al.}{{Sato} et~al.}{2006b}]{gcn5381}
{Sato}, G., et~al. 2006b, GRB Coordinates Network, 5381, 1

\bibitem[\protect\citeauthoryear{{Sato} et~al.}{{Sato} et~al.}{2007b}]{gcn7148}
{Sato}, G., et~al. 2007b, GRB Coordinates Network, 7148, 1

\bibitem[\protect\citeauthoryear{{Savaglio}, {Glazebrook}, \& {Le
  Borgne}}{{Savaglio} et~al.}{2009}]{sgl09}
{Savaglio}, S., {Glazebrook}, K.,  \& {Le Borgne}, D. 2009, \apj, 691, 182

\bibitem[\protect\citeauthoryear{{Schady} et~al.}{{Schady}
  et~al.}{2006}]{gcnr6}
{Schady}, P., et~al. 2006, GCN Report, 6, 1

\bibitem[\protect\citeauthoryear{{Schlafly} \& {Finkbeiner}}{{Schlafly} \&
  {Finkbeiner}}{2011}]{sf11}
{Schlafly}, E.~F.,  \& {Finkbeiner}, D.~P. 2011, \apj, 737, 103

\bibitem[\protect\citeauthoryear{{Schlegel}, {Finkbeiner}, \&
  {Davis}}{{Schlegel} et~al.}{1998}]{sfd98}
{Schlegel}, D.~J., {Finkbeiner}, D.~P.,  \& {Davis}, M. 1998, \apj, 500, 525

\bibitem[\protect\citeauthoryear{{Siegel} et~al.}{{Siegel}
  et~al.}{2010}]{gcnr292}
{Siegel}, M.~H., et~al. 2010, GCN Report, 292, 1

\bibitem[\protect\citeauthoryear{{Soderberg} et~al.}{{Soderberg}
  et~al.}{2006}]{sbk+06}
{Soderberg}, A.~M., et~al. 2006, \apj, 650, 261

\bibitem[\protect\citeauthoryear{{Stamatikos} et~al.}{{Stamatikos}
  et~al.}{2011}]{gcn11553}
{Stamatikos}, M., et~al. 2011, GRB Coordinates Network, 11553, 1

\bibitem[\protect\citeauthoryear{{Stanford} et~al.}{{Stanford}
  et~al.}{2004}]{sdp+04}
{Stanford}, S.~A., {Dickinson}, M., {Postman}, M., {Ferguson}, H.~C., {Lucas},
  R.~A., {Conselice}, C.~J., {Budav{\'a}ri}, T.,  \& {Somerville}, R. 2004,
  \aj, 127, 131

\bibitem[\protect\citeauthoryear{{Stratta} et~al.}{{Stratta}
  et~al.}{2007}]{sdp+07}
{Stratta}, G., et~al. 2007, \aap, 474, 827

\bibitem[\protect\citeauthoryear{{Sullivan} et~al.}{{Sullivan}
  et~al.}{2006}]{slp+06}
{Sullivan}, M., et~al. 2006, \apj, 648, 868

\bibitem[\protect\citeauthoryear{{Suzuki} et~al.}{{Suzuki}
  et~al.}{2010}]{gcn10885}
{Suzuki}, D., {Hayashi}, F., {Kobara}, S., {Sako}, T., {Naito}, H.,  \&
  {Omori}, K. 2010, GRB Coordinates Network, 10885, 1

\bibitem[\protect\citeauthoryear{{Tamura} \& {Ohta}}{{Tamura} \&
  {Ohta}}{2004}]{to04}
{Tamura}, N.,  \& {Ohta}, K. 2004, \mnras, 355, 617

\bibitem[\protect\citeauthoryear{{Uehara} et~al.}{{Uehara}
  et~al.}{2008}]{gcn7223}
{Uehara}, T., et~al. 2008, GRB Coordinates Network, 7223, 1

\bibitem[\protect\citeauthoryear{{Ukwatta} et~al.}{{Ukwatta}
  et~al.}{2008}]{gcnr111}
{Ukwatta}, T.~N., et~al. 2008, GCNR, 111, 1 (2008), 111, 1

\bibitem[\protect\citeauthoryear{{Urata} et~al.}{{Urata}
  et~al.}{2006}]{gcn5717}
{Urata}, Y., et~al. 2006, GRB Coordinates Network, 5717, 1

\bibitem[\protect\citeauthoryear{{van den Bergh}, {Li}, \& {Filippenko}}{{van
  den Bergh} et~al.}{2005}]{vlf05}
{van den Bergh}, S., {Li}, W.,  \& {Filippenko}, A.~V. 2005, \pasp, 117, 773

\bibitem[\protect\citeauthoryear{{Villasenor} et~al.}{{Villasenor}
  et~al.}{2005}]{vlr+05}
{Villasenor}, J.~S., et~al. 2005, \nat, 437, 855

\bibitem[\protect\citeauthoryear{{Wainwright}, {Berger}, \&
  {Penprase}}{{Wainwright} et~al.}{2007a}]{wbp+07}
{Wainwright}, C., {Berger}, E.,  \& {Penprase}, B.~E. 2007a, \apj, 657, 367

\bibitem[\protect\citeauthoryear{{Wainwright}, {Berger}, \&
  {Penprase}}{{Wainwright} et~al.}{2007b}]{wbp07}
{Wainwright}, C., {Berger}, E.,  \& {Penprase}, B.~E. 2007b, \apj, 657, 367

\bibitem[\protect\citeauthoryear{{Wakker}, {Lockman}, \& {Brown}}{{Wakker}
  et~al.}{2011}]{wlb11}
{Wakker}, B.~P., {Lockman}, F.~J.,  \& {Brown}, J.~M. 2011, \apj, 728, 159

\bibitem[\protect\citeauthoryear{{Watson}}{{Watson}}{2011}]{wat11}
{Watson}, D. 2011, \aap, 533, A16

\bibitem[\protect\citeauthoryear{{Xin} et~al.}{{Xin} et~al.}{2011}]{gcn11554}
{Xin}, L.~P., {Zhang}, T.~M., {Qiu}, Y.~L., {Wei}, J.~Y., {Wang}, J., {Deng},
  J.~S., {Wu}, C.,  \& {Han}, X.~H. 2011, GRB Coordinates Network, 11554, 1

\bibitem[\protect\citeauthoryear{{Zheng} \& {Ramirez-Ruiz}}{{Zheng} \&
  {Ramirez-Ruiz}}{2007}]{zr07}
{Zheng}, Z.,  \& {Ramirez-Ruiz}, E. 2007, \apj, 665, 1220

\bibitem[\protect\citeauthoryear{{Ziaeepour} et~al.}{{Ziaeepour}
  et~al.}{2006}]{gcnr21}
{Ziaeepour}, H., et~al. 2006, GCNR, 21, 2 (2006), 21, 2

\bibitem[\protect\citeauthoryear{{Ziaeepour} et~al.}{{Ziaeepour}
  et~al.}{2007}]{gcnr74}
{Ziaeepour}, H., {Barthelmy}, S.~D., {Parsons}, A., {Page}, K.~L., {de
  Pasquale}, M.,  \& {Schady}, P. 2007, GCN Report, 74, 2

\end{thebibliography}


\begin{thebibliography}{0}
\expandafter\ifx\csname natexlab\endcsname\relax\def\natexlab#1{#1}\fi
\expandafter\ifx\csname bibnamefont\endcsname\relax
  \def\bibnamefont#1{#1}\fi
\expandafter\ifx\csname bibfnamefont\endcsname\relax
  \def\bibfnamefont#1{#1}\fi
\expandafter\ifx\csname citenamefont\endcsname\relax
  \def\citenamefont#1{#1}\fi
\expandafter\ifx\csname url\endcsname\relax
  \def\url#1{\texttt{#1}}\fi
\expandafter\ifx\csname urlprefix\endcsname\relax\def\urlprefix{URL }\fi
\providecommand{\bibinfo}[2]{#2}
\providecommand{\eprint}[2][]{\url{#2}}

\end{thebibliography}
\end{document}